\documentclass[10pt,aps,prd,twocolumn,superscriptaddress,longbibliography,nofootinbib]{revtex4-2}

\RequirePackage{snapshot}

\usepackage[utf8]{inputenc}
\usepackage{xcolor}
\usepackage{graphicx}
\usepackage{subfigure}
\usepackage{multirow}
\usepackage{tensor}

\usepackage{amsmath,amssymb,amsfonts}
\usepackage{xcolor}

\usepackage{orcidlink}

\newcommand{\rmd}{{\mathrm d}}

\makeatletter
\newcommand\ID[2]{#1\def\@currentlabel{#1}\label{#2}}
\makeatother

\allowdisplaybreaks

\begin{document}

\title{Formulation Improvements for Critical Collapse Simulations}

\author{Daniela Cors \orcidlink{0000-0002-0520-2600}} 
\affiliation{Friedrich-Schiller-Universität,
	Jena, 07743 Jena, Germany}

\author{Sarah Renkhoff \orcidlink{0000-0002-1233-2593}}
\affiliation{Friedrich-Schiller-Universität,
	Jena, 07743 Jena, Germany}

\author{Hannes R. \surname{Rüter} \orcidlink{0000-0002-3442-5360}}
\affiliation{
  Centro de Astrof\'{\i}sica e Gravita\c c\~ao -- CENTRA,
  Departamento de F\'{\i}sica, Instituto Superior T\'ecnico -- IST,
  Universidade de Lisboa -- UL, Av.\ Rovisco Pais 1, 1049-001 Lisboa,
  Portugal}

\author{David Hilditch \orcidlink{0000-0001-9960-5293}}
\affiliation{
  Centro de Astrof\'{\i}sica e Gravita\c c\~ao -- CENTRA,
  Departamento de F\'{\i}sica, Instituto Superior T\'ecnico -- IST,
  Universidade de Lisboa -- UL, Av.\ Rovisco Pais 1, 1049-001 Lisboa,
  Portugal}

\author{Bernd Brügmann \orcidlink{0000-0003-4623-0525}}
\affiliation{Friedrich-Schiller-Universität,
	Jena, 07743 Jena, Germany}

\begin{abstract}
  The precise tuning required to observe critical phenomena in
  gravitational collapse poses a challenge for most numerical
  codes. First, threshold estimation searches may be obstructed by the
  appearance of coordinate singularities, indicating the need for a
  better gauge choice. Second, the constraint violations to which
  simulations are susceptible may be too large and force searches to
  terminate prematurely. This is a particularly serious issue for
  first order formulations. We want our adaptive pseudospectral code
  \texttt{bamps} to be a robust tool for the study of critical
  phenomena so, having encountered both of these difficulties in work
  on the vacuum setting, we turn here to investigate these issues in
  the classic context of a spherically symmetric massless scalar
  field. We suggest two general improvements. We propose a necessary
  condition for a gauge choice to respect discrete self-similarity
  (DSS). The condition is not restricted to spherical symmetry and
  could be verified with any~$3+1$ formulation. After evaluating
  common gauge choices against this condition, we suggest a
  DSS-compatible gauge source function in generalized harmonic gauge
  (GHG). To control constraint violations, we modify the constraint
  damping parameters of GHG, adapting them to collapse
  spacetimes. This allows us to improve our tuning of the critical
  amplitude for several families of initial data, even going from 6 up
  to 11 digits. This is the most precise tuning achieved with the
  first order GHG formulation to date. Consequently, we are able to
  reproduce the well known critical phenomena as well as competing
  formulations and methods, clearly observing up to 3 echoes.
\end{abstract}

\maketitle

\section{Introduction}
\label{sec:intro}

The threshold of gravitational collapse separates spacetimes at the
verge of black hole formation from those in which a black hole
forms. Spacetimes near this region in solution space have large,
dynamical curvature and, depending on the specific model under
consideration, infinitesimal black holes. This makes them extremely
interesting, albeit hard to treat. In 1993, Choptuik~\cite{Cho93}
tackled the problem by evolving massless scalar fields minimally
coupled to the Einstein field equations in spherical symmetry. He
observed three features bearing a strong resemblance to those observed
near critical points in other fields of physics. He noted that, near
the threshold, the spacetime becomes discretely self-similar (DSS), a
periodic fractal like behavior also referred to as echoing. Second, as
a result of scale invariance, scalars like the black hole
mass~$M_{\text{BH}}$ on one side of the threshold and the Kretschmann
scalar~$I=R_{abcd}{}R^{abcd}$ on the other obey a power law of the
form
\begin{align}
  \label{eq:power_law}
  M_{\text{BH}}\sim |p-p_{\star}|^{\lambda}, \quad
  I_{\text{max}}{}^{-1/4}\sim|p - p_{\star}|^{\lambda} \, ,
\end{align}
as a function to the parametric distance to the threshold, labeled
here by the critical parameter~$p_{\star}$. In the DSS case, this
scaling behavior appears with a superimposed~$\Delta$-periodic
wiggle~\cite{Gun96a,HodPir97}. Third, by evolving different families
of data in spherical symmetry, Choptuik concluded that these features
are universal. In particular, the scaling exponents~$\lambda$ in the
power laws~\eqref{eq:power_law}, as well as the DSS echoing
periods~$\Delta$, were independent of the initial data. Notably, all
the families of initial data, when evolved close enough to the
threshold, share a common configuration, now referred to as the
Choptuik spacetime. A spacetime with this specific symmetry, has since
been proven to exist~\cite{ReiTru19}. These three features
(self-similarity, power law scaling and universality) constitute what
is now known as critical phenomena in gravitational collapse.

Critical phenomena in gravitational collapse have since then been
observed in various matter models, mostly but not exclusively in
spherical symmetry (see~\cite{GunGar07} for a detailed review).
Families of initial data with a varying parameter $p$ are evolved
numerically such that for small values of $p$, the evolution
eventually leads to flat spacetime, whereas large values lead to
horizon formation. Following this end-state classification one can
bisect towards a better estimation of the threshold parameter~$p_*$
delimiting the threshold of collapse. In this way one can measure the
distance to the threshold by the precision with which~$p_*$ is
tuned. The more digits are known, the closer to the threshold the
spacetime is and the more likely we are to observe critical phenomena.

Choptuik achieved an exquisite 13 (or more) digit tuning of~$p_*$ by
using a maximally constrained formulation of general relativity (GR)
with zero shift and areal radius in an adaptive mesh finite
differencing code. Subsequently this setup was studied with a variety
of formulations of GR tailored to spherical symmetry. For instance,
Garfinkle and Duncan~\cite{GarDun98} used null coordinates, and Martín
García and Gundlach~\cite{GarGun03} constructed coordinate systems
adapted to self-similar spherically symmetric spacetimes. There are a
number of studies of the scalar field model with aspherical
perturbations. It was demonstrated numerically that at the linear
level there is only one growing mode solution~\cite{GarGun98}. In
pioneering work~\cite{ChoHirLie03}, non-linear axisymmetric
perturbations of the Choptuik spacetime were studied, and their
findings were confirmed much more recently by Baumgarte
in~\cite{Bau18} with a 13 digit tuning. In the latter, the BSSN
formulation of the Einstein field equations was used with~$1+$log
slicing and $\Gamma$-driver shift condition. Fully 3d non-linear
perturbations have been studied in~\cite{HeaLag13,DepKidSch18},
understandably with far less tuning to the threshold. Of these, Deppe
{et\,al.}'s paper~\cite{DepKidSch18} is of particular interest to us,
because they use a similar setup to ours.  They studied the collapse
of massless scalar fields using the generalized harmonic gauge (GHG)
formulation of GR using a pseudospectral approximation. Coordinates
that gradually zoom into the center of the computational domain and a
new gauge source function were employed. They managed to properly
reproduce Choptuik's results, for the first time with a pseudospectral
code, tuning up to~6 digits.

Choptuik's results are relevant to our understanding of cosmic
censorship, as his spacetime should contain a naked singularity, but
generic physical data are not spherical, and so assessing the
generality of his findings without spherical symmetry is
fundamental. The~6 digit level of tuning of~\cite{DepKidSch18} gives
echoing periods and scaling exponents consistent with those of
spherical symmetry; whereas the finer tuned data
of~\cite{ChoHirLie03,Bau18} find that these scalars are dependent on
the aspherical deviation of the initial data. To assess whether the
different results are due to the difference in tuning, we need to
improve the best level of tuning with GHG and a pseudospectral setup.

Although non-spherical models are generally more difficult to treat,
an understanding is emerging. In
vacuum~\cite{HilWeyBru17,KhiLed18,LedKhi21,SuaRenCor22,BauGunHil23,
  BauBruCor23}, in the presence of electromagnetic
waves~\cite{BauGunHil19,MenBau21} and for scalar fields as mentioned
above, numerical evidence now consistently suggests a deviation from
the spherical phenomenology. Confidence in some of these results, for
instance whether exact DSS occurs beyond spherical symmetry, relies on
the degree of fine tuning to the threshold, as for instance a minimum
number of periods needs to be observed to assess DSS. There is however
room for improvement. For example, in the challenging case of
gravitational waves, the best tuning~\cite{LedKhi21,SuaRenCor22} is
around 6 digits. It should be noted that for one of the two common
families evolved in these works, the oblate centred Brill wave, the
tuning in~\cite{SuaRenCor22} performed with first order GHG was a
digit further from the threshold than the one performed with BSSN
in~\cite{LedKhi21}.

Here we pave the way to improved studies of the threshold of collapse
without symmetry, focusing on a subset of the obstructions to tuning
we found in~\cite{HilWeyBru17,SuaRenCor22} for Brill waves, but
retreating to the classic spherical scalar field setup as a
testbed. In those searches we had to stop tuning either because our
code failed to stably evolve the fields until they dispersed or formed
an horizon, or because our post-processing apparent horizon finder
AHloc3d~\cite{ahloc3d} could not locate the presumably present, but
strangely shaped horizons that formed. Work concerning apparent
horizon finders is ongoing and will be reported on elsewhere. In this
paper we focus instead on solving the problems that prevented our code
from reliably evolving extreme data.

Some families of data were prone to develop undesirable coordinate
features. A shift with a large gradient would be the most common sign
of this instance. Coordinate singularities have been reported to be
the main obstacle of other studies of critical collapse whether
performed with BSSN or GHG. For BSSN evolutions this motivated the use
of ``quasi-maximal'' slicing~\cite{KhiLed18} and of the shock-avoiding
coordinate conditions~\cite{Alc96,Alc02,BauHil22,BauGunHil23}. In GHG,
the remaining gauge freedom lies in the choice of gauge source
functions. The authors of~\cite{DepKidSch18} report that to avoid
divergences, they needed to employ substantially modified gauge source
functions as compared with the standard choice~\cite{LinSzi09} for
binary spacetimes. Our first goal here is therefore to look for a
suitable gauge. Similar in spirit to~\cite{GarGun99}, we look for a
coordinate choice compatible with DSS. In particular we find and check
a necessary condition for compatibility. We then construct a gauge
source function that satisfies this condition.

Large constraint violations were present in the failed evolutions of
most unclassified data. Even using adaptive mesh refinement to help
control error, the closer the spacetimes were to the threshold, the
more constraint violations we observed. For the examples considered
here this is a particularly serious problem, probably since the first
order formulation we employ necessarily introduces a large number of
constraints. Our second goal was therefore to reassess and adjust the
constraint damping parameters of GHG, particularly focusing on the
first order reduction constraints in the context of collapsing
spacetimes. These adjustments make a very significant improvement to
our best tuning and indeed enable our code to display the well known
phenomena first discovered by Choptuik.

In section~\ref{sec:setup}, we present the numerical and physical
setup of our pseudospectral code \texttt{bamps}, which was used to
carry out all the present simulations. In section~\ref{sec:dss} we
derive a DSS-compatibility condition, evaluate most gauge choices
against it and suggest a compatible gauge source function. In
section~\ref{sec:damping} we present a mode analysis and the resulting
improved constraint damping parameters for GHG. Finally, in
section~\ref{sec:cp} we apply these modifications to simulations of
massless scalar fields in spherical symmetry. In
section~\ref{sec:summary} we give a brief summary.

\section{Setup}
\label{sec:setup}
\subsection{Code \texttt{bamps}}
\label{subsec:bamps}
\begin{figure}[t]
\includegraphics[width=0.45 \textwidth]{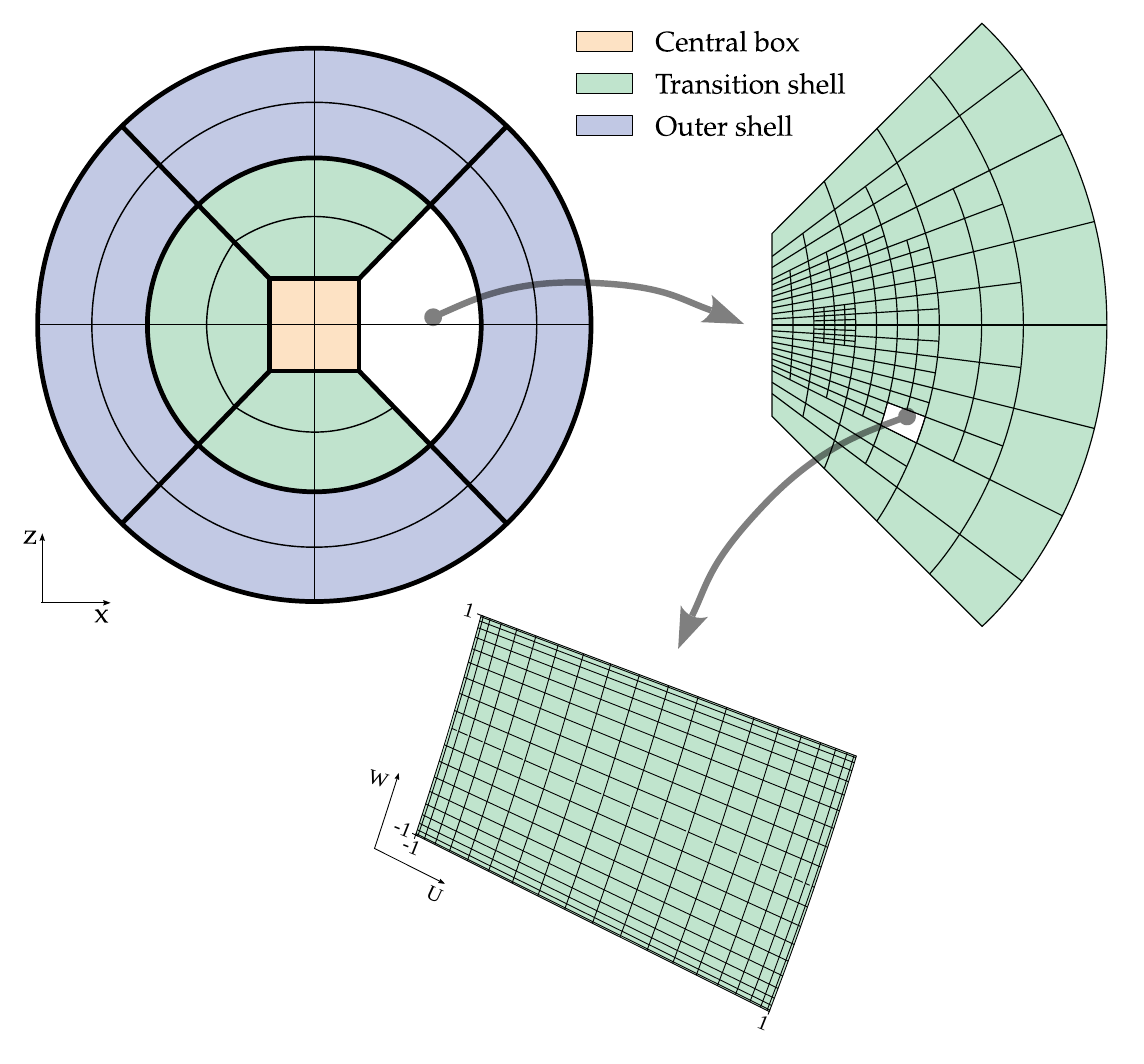}
\caption{The \texttt{bamps} grid structure illustrating h-refinement.}
\label{fig:cubedball}
\end{figure}

All the simulations we present in this paper were carried out with the
pseudospectral code \texttt{bamps}~\cite{HilWeyBru15}. Its grid is
divided into three main patches: a cubed box at the center of the
simulation domain, a transition shell in the shape of a cubed sphere,
and a spherical shell circling the domain up to the outer boundary
(see Fig.~\ref{fig:cubedball}). Each of these patches is itself
divided into cells, inside of which the evolved fields are stored on
Chebyshev-Gauss-Lobatto collocation points.

hp-adaptive mesh refinement (AMR) controls the resolution that every
portion of the fields is granted by locally adapting the number of
collocation points. It does so both by adding or removing points in a
given cell (p-refinement), and by subdividing or joining cells
(h-refinement). In order to evaluate the refinement needed, AMR can
assess the error produced during the simulation and also the
smoothness of the evolved fields. Details of the refinement scheme are
given in~\cite{RenCorHil23}.

Neighboring cells communicate with each other through characteristic
fields using the penalty method. At the outer boundary, we use the
constraint preserving boundary conditions described in~\cite{Rin06a}
and~\cite{HilWeyBru15}. Symmetries are handled by the cartoon method,
using double cartoon for spherically symmetric runs.

In \texttt{bamps} we employ a standard free evolution approach. The
constraints are only explicitly solved for the initial data but not at
any timestep during the evolution. The scalar field initial data is
generated by an elliptic solve integrated within \texttt{bamps}, which
uses the hyperbolic relaxation method~\cite{RueHilBug17}.

Time evolution is carried out using a fourth order Runge-Kutta
method. The timesteps are adjusted during the evolution in order to
preserve the Courant-Friedrich-Lewy (CFL) condition.

\subsection{Physical System}
\label{subsec:GR}

In geometric units the trace reversed Einstein field equations read
\begin{align}
  \label{eq:EFE}
  R_{ab} = 8\pi \left(T_{ab} -\frac{1}{2}g_{ab} T\right) \, ,
\end{align}
where~$g_{ab}$ is the~$4$d metric, $R_{ab}$ the Ricci tensor, $T_{ab}$
the energy momentum tensor and~$T=g^{ab}T_{ab}$ its trace. Time
evolutions are enabled by a~$3+1$ splitting of the~$4$d
metric~$g_{ab}$. This gives the~$3$d spatial metric $\gamma_{ij}$ and
a normal unit vector~$n^a=\alpha^{-1}(1, -\beta^i)$, where~$\alpha$ is
the lapse and~$\beta^i$ the shift. In these variables the line element
becomes
\begin{align}
  \label{eq:3+1}
  ds^2 = -\alpha^2 \,dt^2
  + \gamma_{ij}\,(\beta^i dt + dx^i)(\beta^j dt + dx^j) \,.
\end{align}
We denote~$4$d component indices with Latin letters starting from $a$
and~$3$d spatial ones starting from~$i$.
   	  
We consider a massless real scalar field~$\varphi$ minimally coupled
to the Einstein field equations. The corresponding scalar field energy
momentum tensor is given by
\begin{align}
  \label{eq:energy_momentum_tensor}
  T_{ab} = \nabla_a \varphi \nabla_b \varphi - \frac{1}{2} g_{ab}(
  \nabla^c \varphi \nabla_c \varphi)\,.
\end{align}

\subsection{Initial Data}
\label{subsec:initialdata}

In this work we consider two families of initial data, one starting at
a moment of time symmetry, the other with a predominantly incoming
pulse. The moment of time symmetry data is constructed from a scalar
field with a Gaussian profile of the form
\begin{align}
  \label{eq:initial_phi_mts}
  \varphi = A \left( e^{-(r + R_0)^2} + e^{-(r - R_0)^2} \right)\,,
\end{align}
and with a vanishing derivative along~$n^a$
\begin{align}
  \label{eq:initial_pi_mts}
  n^a \nabla_a \varphi = 0\, .
\end{align}
For moment of time symmetry initial data we use~$R_0 = 0$ as
in~\cite{Bau18}. Observe that~\eqref{eq:initial_phi_mts} implies
that our $A$ is a half of the equivalent parameter $\eta$ used
in~\cite{Bau18}.
               
For the incoming initial data we use the same initial scalar field as
in~\cite{DepKidSch18}. Thus we have
\begin{align}
  \label{eq:initial_phi_inc}
  \varphi = A \, \frac{e^{-(r + R_0)^2} - e^{-(r - R_0)^2}}{r}
\end{align}
and
\begin{align}
  \label{eq:initial_pi_inc}
  \begin{split}
    n^a \nabla_a \varphi & =    \\
    & 2 A \frac{ (r + R_0) e^{-(r + R_0)^2} + (r - R_0) e^{-(r -
        R_0)^2}}{r}\,,
  \end{split}
\end{align}
with~$R_0 = 5$ and the appropriate limits taken at the origin.

To solve the ADM constraints we make a conformal decomposition of the
metric and solve the extended conformal thin sandwich equations
(XCTS)~\cite{BauSha10,Tic17}.  We choose the conformal metric to be
flat, $\bar \gamma_{ij} = \delta_{ij}$, and take its time derivative
to vanish, $\partial_t \bar \gamma_{ij} = 0$. Furthermore we impose
maximal slicing on the initial data, so that both the trace of the
extrinsic curvature and its time derivative vanish,
$K=\gamma^{ij}K_{ij} = 0$, $\partial_t K = 0$.  After fixing these
variables the XCTS equations become a set of coupled elliptic partial
differential equations for the conformal factor~$\psi$, the
shift~$\beta^i$ and the lapse~$\alpha$.

At the outer boundary, $\partial \Omega$, we impose Robin boundary
conditions compatible with a~$1/r$ decay for~$\psi$, $\beta^i$
and~$\alpha$. Concretely, the boundary conditions are given by
\begin{align}
	\label{eq:initial_data_boundary_psi}
	s^i \partial_i \psi |_{\partial \Omega} 
	&= \frac{1 - \psi}{r}\, , \\
	\label{eq:initial_data_boundary_alpha}
	s^i \partial_i \alpha |_{\partial \Omega} 
	&= \frac{1 - \alpha}{r}\, , \\
	\label{eq:initial_data_boundary_beta}
	s^i \partial_i \beta^j |_{\partial \Omega} 
	&= \frac{\beta^j}{r} \, ,
\end{align}
where~$s^i = \bar \gamma^{ij} \partial_j r / L$ is the spatial normal to
the outer boundary, with the normalization factor~$L$ such
that~$\bar \gamma^{ij} s_i s_j=1$.

For incoming initial data we have to solve the full coupled set of the
XCTS equations. For moment of time symmetry initial data on the other
hand, we only have to solve one equation for~$\psi$. Thanks to the
choice in Eq.~\eqref{eq:initial_pi_mts} the solution for shift and
lapse are trivially given by~$\beta^i = 0$ and~$\alpha = 1$. The
remaining equation for~$\psi$ has the form
\begin{align}
  \label{eq:initial_data_pde_gaussian}
  0 = \delta^{ij} \partial_i \partial_j \psi
  + \pi \psi \delta^{ij} \partial_i \varphi \partial_j \varphi\, ,
\end{align}
where~$\pi$ here is the mathematical constant not to be confused with
the field $\pi$ introduced later. We use the hyperbolic relaxation
method~\cite{RueHilBug17} implemented in \texttt{bamps} to solve the
XCTS equations or Eq.~\eqref{eq:initial_data_pde_gaussian}.

\subsection{Evolution Equations}
\label{subsec:evolution}
\paragraph{Scalar field evolution equations}
\label{subsubsec:sf_evolution}

The equation of motion of a massless scalar field is given by the
Klein Gordon equation~$\nabla^a \nabla_a \varphi = 0$. We work under a
first order reduction with~$\pi$ as the time reduction
variable~$n^a \partial_a \varphi$, and~$\chi_i$ as the spatial
reduction variable associated to the reduction
constraint~$S_i := \partial_i \varphi - \chi_i = 0$. This yields the
first order massless Klein Gordon evolution system
\begin{align}
  \label{eq:sf_rhs}
  \partial_t \varphi
  &=\alpha \pi + \beta^i \chi_i \, ,\\
  \partial_t \pi
  &=\beta^i \partial_i \pi + \gamma^{ij}
    (\chi_j \partial_i \alpha + \alpha \partial_i \chi_j
    - \alpha {}^{(3)}\Gamma^k{}_{ij} \chi_k) \\
  & \quad + \alpha \pi K + \sigma \beta^i S_i \, ,\nonumber\\*
  \partial_t \chi_i
  &= \pi \partial_i \alpha + \alpha \partial_i \pi
    + \chi_j \partial_i \beta^j + \beta^j \partial_j \chi_i
    + \sigma \alpha S_i \, ,
\end{align}
where~$\sigma$ is a damping term and the spatial connection
is denoted by~${}^{(3)}\Gamma^i{}_{jk}$.

\paragraph{GHG evolution equations}
\label{subsubsec:ghg}

In~\texttt{bamps}, the Einstein field equations are formulated with
the first order reduction of the GHG
formulation~\cite{LinSchKid05}. The evolved variables are the metric
components~$g_{ab}$, the time reduction variable~$\Pi_{ab}$
corresponding to~$-n^d \partial_d g_{ab}$ and the spatial reduction
variable~$\Phi_{iab}$ associated to the reduction
constraint~$C_{iab}:=\partial_ig_{ab}-\Phi_{iab} = 0$. The evolution
equations are
\begin{align}
  \label{eq:ghg_g}
  \partial_t \tensor{g}{_a_b}
  &=\beta^i \partial_i \tensor{g}{_a_b}
    - \alpha \tensor{\Pi}{_a_b}
    + \gamma_1 \beta^i \tensor{C}{_i_a_b} \, , \\
  \partial_t \tensor{\Pi}{_a_b}
  &=\beta^i \partial_i \tensor{\Pi}{_a_b}
    - \alpha \gamma^{ij} \partial_i \tensor{\Phi}{_j_a_b}
    + \gamma_1 \gamma_2 \beta^i \tensor{C}{_i_a_b} \nonumber\\*
  & + 2 \alpha g^{cd}
    \left( \gamma^{ij} \tensor{\Phi}{_i_c_a} \tensor{\Phi}{_j_d_b}
    - \tensor{\Pi}{_c_a} \tensor{\Pi}{_d_b}
    - g^{ef} \tensor{\Gamma}{_a_c_e} \tensor{\Gamma}{_b_d_f} \right) \nonumber\\*
  & - 2 \alpha \left( \tensor{\nabla}{_(_a} \tensor{H}{_b_)}
    + \gamma_4 \tensor{\Gamma}{^c_a_b} \tensor{C}{_c}
    - \frac{1}{2} \gamma_5 \tensor{g}{_a_b} \Gamma^c \tensor{C}{_c} \right) \nonumber\\*
  &-\frac{1}{2} \alpha n^c n^d \tensor{\Pi}{_c_d} \tensor{\Pi}{_a_b}
    - \alpha n^c \gamma^{ij} \tensor{\Pi}{_c_i} \tensor{\Phi}{_j_a_b} \nonumber\\*
  & + \alpha \gamma_0 \left( 2 \tensor{\delta}{^c_{(a}} \tensor{n}{_{b)}}
    - \tensor{g}{_a_b} n^c \right) \tensor{C}{_c}\nonumber\\*
  & - 16 \pi \alpha 
    \left (T_{ab} - \frac{1}{2} g_{ab} \tensor{T }{^c_c} \right) \, ,
    \label{eq:ghg_Pi}\\
  \partial_t \tensor{\Phi}{_i_a_b}
  &= \beta^j \partial_j \tensor{\Phi}{_i_a_b}
    - \alpha \partial_i \tensor{\Pi}{_a_b}
    + \gamma_2 \alpha \tensor{C}{_i_a_b} \nonumber\\*
  & + \frac{1}{2} \alpha n^c n^d
    \tensor{\Phi}{_i_c_d} \tensor{\Pi}{_a_b}
    + \alpha\gamma^{jk} n^c \tensor{\Phi}{_i_j_c}
    \tensor{\Phi}{_k_a_b}\,.\label{eq:ghg_Phi}
\end{align}
The canonical constraint damping parameters are~$\gamma_1 = -1$
and~$\gamma_4 = \gamma_5 = 1/2$ as introduced in~\cite{HilWeyBru15}.
We use~$\alpha\gamma_0=2$ for incoming data and~$\alpha\gamma_0=4$ for
the moment of time symmetry data. In section~\ref{sec:damping} we
examine the role of~$\gamma_2$, directly analogous to the damping
parameter of the scalar field~$\sigma$, in relation to the spatial
reduction constraint.  The harmonic constraint corresponds to
\begin{align}
	\label{eq:harmonic_constraint}
	C_a:= H_a + \Gamma_a = 0 \, ,
\end{align}
where~$H_a$ is the gauge source function, which we are free to specify
(see section~\ref{subsec:gsf}). As pointed out in~\cite{LinSchKid05},
$\lbrace C_a,\partial_t C_a\rbrace = \lbrace 0,0\rbrace$ encodes the
Hamiltonian and momentum constraints, thereby ensuring both that the
coordinates are harmonic and that the Einstein field equations are
satisfied. The system of equations~\eqref{eq:ghg_g}-\eqref{eq:ghg_Phi}
is symmetric hyperbolic, ensuring a well-posed initial boundary value
problem provided suitable initial and boundary values.

\subsection{Gauge source functions}
\label{subsec:gsf}

Harmonic coordinates satisfy
\begin{align}
	\label{eq:harmonic_coord}
	\Box x^a &= - \Gamma^a = 0\, .
\end{align}
The GHG formulation generalizes this gauge to a less restrictive
condition with the addition of a source term~$H_a$ to the right hand
side of~\eqref{eq:harmonic_coord}. This generalization is satisfied as
long as the harmonic constraint~\eqref{eq:harmonic_constraint},
$C_a = 0$, is satisfied. All the gauge freedom of GHG is contained
in~$H_a$. When the harmonic constraint is satisfied, the expressions
of~$\Gamma^a$ in standard~$3+1$ formalism give the following evolution
equations for the lapse and shift
\begin{align}
  \label{eq:dt_alpha_ghg}
  \rmd_t\alpha
  &= -\alpha^2 \big(n^a H_a + K\big) \, , \\
  \label{eq:dt_beta_ghg}
  \rmd_t\beta^i
  &= \alpha^2 \big(H^i +{}^{(3)}\Gamma^i - \partial^i\ln\alpha\big)\, ,
\end{align}
where~$\rmd_t=\alpha \,n^a \partial_a=\partial_t -\beta^k\partial_k$ and
the~$3$d contracted spatial connection
is~${}^{(3)}\Gamma^i={}^{(3)}\Gamma^i{}_{jk}\gamma^{jk}$.

To guarantee symmetric hyperbolicity, the arbitrary function of
spacetime~$H_a$ cannot contain derivatives of the fields. In that
sense the gauge freedom of GHG is rather limited. We assess here in
detail the choice of Harmonic Damped Wave Gauge (HDWG) as well as two
modifications of it used in studies of critical collapse.

\paragraph{HDWG}\label{subsubsec:hdwg}

As explained in~\cite{LinSzi09}, the harmonic
condition~\eqref{eq:harmonic_coord} can lead to strong gauge dynamics,
when the physical degrees of freedom are very dynamical. An approach
to combat this effect is to reduce those gauge dynamics, with the time
and spatial coordinates subject to slightly different considerations.
For instance, to suppress gauge dynamics associated with the spatial
coordinates, the shift can directly be damped through the gauge source
function yielding a shift condition very similar to
the~$\Gamma$-driver one~\eqref{eq:gamma_driver}. The equations for
harmonic slicing and a shift damped by a factor~$\eta$ in GHG are
\begin{align}
  \label{eq:hzero}
  H_a &= -\eta\, \gamma_{ai}\beta^i \alpha^{-1} \, ,\\
  \rmd_t\alpha &= -\alpha^2 K \, ,\\
  \rmd_t\beta^i &= -\alpha\eta\beta^i +\alpha^2 \big({}^{(3)}\Gamma^i
                  - \partial^i\ln\alpha\big)\,.
\end{align}
This choice of~$H_a$ only damps the spatial gauge dynamics, but an
extra term is necessary in order to damp the temporal gauge dynamics.
Besides, the authors of~\cite{LinSzi09} report a concern about the
growth of $\alpha^{-1}\sqrt{\gamma}$ in some simulations,
where~$\gamma=\det(\gamma_{ij})$ is the spatial volume element.
Since~$n^a H_a =n^a \partial_a \ln(\alpha^{-1}\sqrt{\gamma}) -
\alpha^{-1}\partial_k \beta^k$, they suggest to
fix~$n^a H_a=-\eta_L\,\ln(\alpha^{-1}\sqrt{\gamma})$ in order to
exponentially suppress~$\alpha^{-1}\sqrt{\gamma}$. This is the well
known HDWG for GHG (see equation (A15) in~\cite{LinSzi09})
\begin{align}
  \label{eq:hdwg}
  H_a &= \eta_L \,\ln(\alpha^{-1}\sqrt{\gamma})n_a
        -\eta_S \,\gamma_{ai}\beta^i \alpha^{-1} \, ,\\
  \label{eq:hdwg_dtalpha}
  \rmd_t\alpha &= \alpha^2 \big(\,\eta_L\ln(\alpha^{-1}\sqrt{\gamma})
                 - K\big) \, ,\\
  \rmd_t\beta^i &= -\alpha\eta_S\beta^i +\alpha^2 \big({}^{(3)}\Gamma^i
                  - \partial^i\ln\alpha\big) \, ,
\end{align}
with a canonical choice of coefficients being~$\eta_L=1$
and~$\eta_S=2$. Although the HDWG choice of~$H_a$ was extremely
successful yielding black hole simulations for the authors
of~\cite{SziLinSch09}, in critical collapse studies, we have found
that it fell short to handle the spacetimes of interest. Some
modifications of this function have shown to be more appropriate for
critical collapse studies, as we present below.

\paragraph{HDWG-$\alpha^2$ $H_a$}
\label{subsubsec:a2_gsf}

In~\cite{HilWeyBru17}, equations (1) and (5), and more recently
in~\cite{SuaRenCor22}, equation (4), we gave a variant of the HDWG
gauge source function characterised by the powers of the lapse as
\begin{align}
  \label{eq:a2_gsf}
  H_a &= \eta_L\alpha^{-2} \,\ln(\alpha^{-1}\sqrt{\gamma})n_a
        -\eta_S \,\gamma_{ai}\beta^i \alpha^{-2}\, ,\\
  \label{eq:a2_dtalpha}
  \rmd_t\alpha &= \,\eta_L\ln(\alpha^{-1}\sqrt{\gamma}) - \alpha^2 K \, ,\\
  \rmd_t\beta^i &= -\eta_S\beta^i +\alpha^2 \big({}^{(3)}\Gamma^i
                  - \partial^i\ln\alpha\big)\,.
\end{align}
These modifications take into account the collapse of the lapse in
spacetimes of extreme curvature (see a description in
section~\ref{sec:damping}) so that relevant terms survive despite a
vanishing lapse. In the shift evolution equation one can see that the
damping of the spatial gauge dynamics takes place whether or not the
lapse collapses. This success inspired one of the modifications we
present below, the adjusted damping parameter
(see~\ref{sec:damping}). Large values of~$\eta_S$ had to be used to
deal with undesirable coordinate features for some families of data
in~\cite{SuaRenCor22}. Empirically we found~$\eta_L=1$ and~$\eta_S=2$
to be a reliable choice.

\paragraph{HDWG-$\ln(\alpha)$ $H_a$}
\label{subsubsec:ln_gsf}

In~\cite{DepKidSch18} the authors introduce a new version of the HDWG
gauge source function in order to further enhance the suppression
of~$\alpha^{-1}\sqrt{\gamma}$ and to control the growth
of~$\alpha^{-1}$.
\begin{align}
  \label{eq:ln_gsf}
  \begin{split}
  H_a = {}& R(t) W(x^i) \\
       & \Bigg( \Big (
        \big(\ln(\alpha^{-1}\sqrt{\gamma})\big)^5
        + \big(\ln(\alpha^{-1})\big)^5\Big)n_a\\
      & \phantom{\Bigg(}
        - \big(\ln(\alpha^{-1}\sqrt{\gamma})\big)^4
        \gamma_{ai}\beta^i \alpha^{-1} \Bigg)\, ,
  \end{split}\\
  \label{eq:ln_dtalpha}
  \begin{split}
  \rmd_t\alpha
      ={}& R(t) W(x^i) \alpha^2 \\
       &  \left (\big(\ln(\alpha^{-1}\sqrt{\gamma})\big)^5
        +\big(\ln(\alpha^{-1})\big)^5 -  K\right)\, ,
  \end{split}\\
  \begin{split}
  \rmd_t\beta^i
      ={}& R(t) W(x^i) \\
       & \left( -\alpha \big(\ln(\alpha^{-1})\big)^4 \beta^i \right)
        +\alpha^2 \big( {}^{(3)}\Gamma^i - \partial^i\ln\alpha\big) \,,
  \end{split}
\end{align}
where~$R(t)$ is a roll-on function and~$W(x^i)$ is a spatial weight
function. $R(t)$ allows to start simulations with maximal slicing and
then move on smoothly to their modified HDWG whilst~$W(x^i)$ enforces
pure harmonic gauge at the boundary. As we are only interested in
testing the impact of the modified HDWG, in our tests we
consider~$R(t)=W(x^i)=1$. The authors of~\cite{DepKidSch18} also
mention that for reliable evolutions of critical collapse, the powers
in these logarithms are higher than the ones typically needed in
binary black hole evolutions.

\section{DSS-compatible coordinates} 
\label{sec:dss}

As is hinted in the previous subsection, choosing a gauge for critical
collapse simulations is not trivial. In GHG, gauge source functions
have been constructed by careful experimentation and used to control
divergences and a collapsing lapse. Even then, code failures
coinciding with undesirable coordinate features such as a shift with a
large gradient, occur. Ultimately this may result in a coordinate
singularity. In practice, since these coordinate features trigger
large constraint violations they tend to result in a crash before
unambiguously determining the presence of such a
singularity. Nevertheless, our best hypothesis thus far is that at
least a subset of our code crashes are due to these undesirable
coordinate features. Therefore we look for a gauge choice that keeps
the lapse and shift ``well behaved'' during the simulation by reducing
any additional coordinate features.

A natural suggestion is to tailor coordinates to the symmetries of the
studied system. In the case of DSS spacetimes, using DSS-adapted
coordinates would have the key advantage of directly tackling the
problem of coordinate singularities since, if one could compute
through one complete period, subsequent periods ought to follow
without extra gauge features, as long as numerical error were
sufficiently well controlled. From the numerical point of view such
adapted coordinates are therefore attractive because they should
potentially reduce the computational cost to reach a desired
error. In~\cite{GarGun99}, the authors study coordinates based on the
ADM equations and on the homothetic Killing vector describing the
continuous self-similarity (CSS) or DSS present in critical collapse
spacetimes. With a particular boundary prescription, they test two
gauges adapted to self-similarity in spherical symmetry. They call
these coordinates \emph{symmetry seeking}. Unfortunately neither is a
simple modification of popular dynamical gauge conditions now known to
be well behaved when treating a range of generic~3+1 dimensional data.
Therefore, with the same general motivation, we take a different
approach. Starting from the expression of the assumed symmetry---DSS
in this case---on the evolved variables, we derive a necessary
condition for gauge choices to be \emph{compatible} with DSS (CSS
would be similar). This condition is not restricted to spherical
symmetry and can be straightforwardly used with any~3+1 formulation of
GR.

\subsection{DSS-compatibility condition}
\label{subsec:DSS_cond}
\paragraph{Periodic rescaling}
\label{subsubsec:periodic_rescaling}

By definition~\cite{GunGar07}, there exist coordinates~$(T, x^i)$ in
which the metric of a DSS spacetime displays a periodic conformal
rescaling of the form
\begin{align}
  \label{eq:DSS_rescale}
  g_{ab} \left(T, x^i\right)
  = e^{-2T} \tilde{g}_{ab} \left(T, x^i\right) \, ,
\end{align}
where the conformal metric~$\tilde{g}_{ab}$ is periodic, with fixed
period~$\Delta$
\begin{align}
  \label{eq:DSS_period}
  \tilde{g}_{ab} \left(T + \Delta, x^i\right)=
  \tilde{g}_{ab} \left(T, x^i \right) \, .
\end{align}
Such coordinates are referred to as \emph{DSS-adapted}
coordinates. Slow time is a time coordinate adapted to DSS. It can be
thought of as the logarithm of an ever decreasing spacetime scale. An
example of slow time is the one computed as the logarithmic distance
of proper time~$\tau$ to the accumulation time~$\tau_{*}$
\begin{align}
  \label{eq:slow_time_prop}
  T_p=-\ln|\tau_{*} -\tau|\, ,
\end{align}
along a future directed timelike curve that terminates at the
accumulation point. (In general the coordinate then needs to be
suitably extended away from the observer). In~$(T, x^i)$ coordinates,
the characteristic time scale is~$\Delta$, therefore it would be
numerically beneficial to evolve the spacetime using~$T$, avoiding
then the need to resolve ever decreasing time scales.

Obviously, combining~\eqref{eq:DSS_rescale} and~\eqref{eq:DSS_period}
we obtain that, after a period~$\Delta$, the~$4$d metric is rescaled
as
\begin{align}
  \label{eq:DSS_4d}
  g_{ab} \left(T + \Delta, x^i\right)
  =  e^{-2\Delta} g_{ab} \left(T, x^i\right) \, .
\end{align}
This transformation of the metric can be translated to the~3+1
variables in~\eqref{eq:3+1} as
\begin{align}
  \label{eq:DSS_3+1_1}
  \gamma_{ij} \left(T + \Delta, x^k\right)
  &= e^{-2\Delta}\gamma_{ij} \left(T, x^k\right) \, ,\\
\label{eq:DSS_alpha}
  \alpha \left(T + \Delta, x^i\right)
  &= e^{-\Delta}\alpha \left(T, x^i\right) \, ,\\
\label{eq:DSS_beta}
  \beta^{i} \left(T + \Delta, x^j\right)
  &= \beta^{i} \left(T, x^j\right)\, .
\end{align}
The transformation after a period~$\Delta$ of other relevant
quantities can be derived from these, giving
\begin{align}
  \sqrt{\gamma\left(T + \Delta, x^i\right) }
  &= e^{-3\Delta} \sqrt{\gamma \left(T, x^i\right)}\, ,\\
  {}^{(3)}\Gamma^i \left(T + \Delta, x^j\right)
  &= e^{2\Delta}\,{}^{(3)}\Gamma^i \left(T, x^j\right) \, ,\\
\label{eq:DSS_3+1_2}
K \left(T + \Delta, x^i\right) &=e^{\Delta} K \left(T, x^i\right)\, .
\end{align}

\paragraph{Coordinate choice}
\label{subsubsec:coord_choice}

Let us briefly refer to~\eqref{eq:DSS_3+1_1}-\eqref{eq:DSS_3+1_2}
holding at all times as (i). A consequence of these, in particular
of~\eqref{eq:DSS_alpha}-\eqref{eq:DSS_beta}, is that (ii)
\begin{align} 
	\label{eq:DSS_condition_dTlapse}
	\rmd_T \alpha (T + \Delta, x^i)
	&=  e^{-\Delta} \rmd_T \alpha (T, x^i) \, , \\
	\label{eq:DSS_condition_dTshift}
	\rmd_T \beta^i (T + \Delta, x^j)
	&= \rmd_T \beta^i (T, x^j) 
\end{align}
hold at all times too. Thus (ii) is necessary for (i), and in turn for
coordinates to be adapted to DSS. It is not a sufficient condition.
(ii) does not imply (i) because the time integration required in that
step is also affected by the remaining degrees of freedom of GR.

In the $3+1$ split, we choose coordinates by imposing evolution
equations for the lapse and shift
\begin{align}
	\label{eq:general_slicing_cond}
	\rmd_t \alpha &= F[g_{ab}] \, , \\
	\label{eq:general_shift_cond}
	\rmd_t \beta^i &= G^i[g_{ab}] \, ,
\end{align}
where the right hand-sides are functionals permitted to depend also upon
derivatives of the metric. Given this gauge choice, (ii) implies that (iii)
\begin{align}
	\label{eq:DSS_condition_lapse}
	F \left[g_{ab} \left(T + \Delta, x^i\right)\right]
	&= e^{-\Delta} F \left[g_{ab} \left(T, x^i\right)\right] \,, \\
	\label{eq:DSS_condition_shift}
	G^i \left[g_{ab} \left(T + \Delta, x^j\right)\right]
	&= G^i \left[g_{ab} \left(T , x^j\right)\right]
\end{align}
holds, with the arguments assumed to respect such a symmetry according
to~\eqref{eq:DSS_3+1_1}-\eqref{eq:DSS_beta}, namely (i) holds. (iii)
is a necessary condition for (ii), because without (iii) it cannot be
the case that (ii) holds for the given gauge choice. By transitivity,
(iii) is also a necessary condition of (i). We call (iii) the
\emph{DSS-compatibility} necessary condition for an
arbitrary choice of gauge evolution equations to render the associated
coordinates DSS-adapted.

Since (ii) is not sufficient to have (i) DSS-adapted coordinates,
neither is (iii). Note that (iii) is not even a sufficient condition
for (ii) to hold, as without additionally assuming (i), (iii) alone
could not yield (ii)
\begin{align*}
  &(i) \Rightarrow (ii) \Rightarrow (iii) \, , \\*
  &(iii) \not \Rightarrow (ii) \not \Rightarrow (i)  \, .
\end{align*}

One might hope that satisfying the DSS-compatible condition would at
least remove undesirable gauge features emerging from the
incompatibility between the gauge choice and the self-similarity. This
in itself would also be a good step towards avoiding coordinate
singularities, the best step being of course actually evolving in
adapted coordinates as explained at the beginning of this section.

Evolving a spacetime in DSS-adapted coordinates is not needed to
assess the presence of exact DSS, as this can be verified after the
evolution through coordinate transformations from the evolved
coordinates to constructed DSS-adapted ones (see for instance
section~\ref{subsec:DSS_results}). However, it would not only be
desirable from the numerical point of view (potential efficiency and
coordinate singularity avoidance), but also because the dynamical
construction of DSS-adapted coordinates would be the least ambiguous
demonstration of such a symmetry.

In practice we can now use
conditions~\eqref{eq:DSS_condition_lapse}-\eqref{eq:DSS_condition_shift}
to assess whether the different gauges that are commonly used in
critical collapse studies may be compatible with the symmetry we
expect at the threshold of collapse. We refer to gauge choices
satisfying the DSS-compatibility
condition~\eqref{eq:DSS_condition_lapse}
and~\eqref{eq:DSS_condition_shift} as \emph{DSS-compatible}. If a
gauge is not DSS-compatible, the associated coordinates will not be
DSS-adapted.

\subsection{Moving puncture gauge}
\label{subsec:no_dss_gauge_z4}
\paragraph{Bona-Masso slicing condition}
\label{subsubsec:BMslicing}

Most finite difference simulations using the
BSSN~\cite{BauSha98,ShiNak95,NakOohKoj87} or conformal Z4
formulations~\cite{BerHil09,AliBonBon11,HilBerThi12} choose the
Bona-Masso~\cite{BonMasSei94} family of slicing conditions
\begin{align}
  \rmd_t \alpha = - \alpha^2 f(\alpha) K \, .
\end{align}
Harmonic gauge corresponds to~$f(\alpha)=1$. The use
of~\eqref{eq:DSS_alpha} and~\eqref{eq:DSS_3+1_2} gives
\begin{align}
  \rmd_t \alpha \left(T + \Delta, x^i\right)
  &= - e^{-\Delta} \alpha^2 \left(T, x^i\right) K \left(T, x^i\right)\\
  &=  e^{-\Delta}\rmd_t \alpha \left(T, x^i\right) \, ,
\end{align}
which satisfies the DSS-compatibility
condition~\eqref{eq:DSS_condition_lapse}. More interesting is to
consider~$1+\log$ slicing, which corresponds
to~$f(\alpha)=\frac{2}{\alpha}$, as it is the most popular
choice. After a period~$\Delta$, we obtain
\begin{align}
  \rmd_t \alpha \left(T + \Delta, x^i\right)
  &= -2 \alpha \left(T, x^i\right) K \left(T, x^i\right)\nonumber\\*
  &\neq e^{-\Delta} \rmd_t \alpha \left(T, x^i\right) \,  ,
\end{align}
which does not satisfy the
condition~\eqref{eq:DSS_condition_lapse}. Similarly, shock avoiding
slicing, corresponding to~$f(\alpha)=1 + \frac{\kappa}{\alpha^2}$ for
a given constant~$\kappa$, also fails to satisfy the DSS-compatibility
condition
\begin{align}
  \rmd_t \alpha \left(T + \Delta, x^i\right)
  &= \left(- e^{-\Delta} \alpha^2 \left(T, x^i\right)
    + e^{+\Delta}\kappa\right) K \left(T, x^i\right)\nonumber\\*
  &\neq e^{-\Delta} \rmd_t \alpha \left(T, x^i\right)\, .
\end{align}

\paragraph{$\Gamma$-Driver shift condition}
\label{subsubsec:GDshift}

The conformal nature of the BSSN and conformal Z4 formulations plays a
role in the choice of shift condition.  These formalisms are expressed
in terms of a conformal spatial
metric~$\bar{\gamma}_{ij}= \chi \gamma_{ij}$ where the conformal
factor is proportional to the determinant of the spatial
metric~$\chi= \gamma^{-1/3}$. The Christoffel symbols associated to
the conformal spatial metric~$\bar{\Gamma}^i{}_{jk}$ can be contracted
with this metric to
give~${}^{(3)}\bar{\Gamma}^i = \bar{\gamma}^{jk}
{}^{(3)}\bar{\Gamma}^i{}_{jk}$.
Using~\eqref{eq:DSS_3+1_1}-\eqref{eq:DSS_3+1_2} we obtain
\begin{align}
  {}^{(3)}\bar{\Gamma}^i \left(T + \Delta, x^j\right)
  = {}^{(3)}\bar{\Gamma}^i \left(T, x^j\right).
\end{align}

In terms of this connection, the~$\Gamma$-driver condition reads as
\begin{align}
  \label{eq:gamma_driver}
  \rmd_t \beta^i = -\mu_S {}^{(3)}\bar{\Gamma}^i - \eta \beta^i\,,
\end{align}
with~$\mu_S$ a constant. We find that this choice does satisfy the
DSS-compatibility condition~\eqref{eq:DSS_condition_shift}
\begin{align}
  \rmd_t \beta^i \left(T + \Delta, x^i \right)
  &=\rmd_t \beta^i \left(T , x^i \right) \, .
\end{align}
We conclude that the most common gauge choices used in finite
differencing codes will not provide coordinates adapted to DSS due to
the slicing condition, with the exception of pure harmonic gauge.

\subsection{Gauge in GHG} 
\label{subsec:no_dss_gauge_GHG}

Since in GHG the gauge freedom is fixed through the gauge source
function~$H_a$, the DSS-compatibility
condition~\eqref{eq:DSS_condition_lapse}-\eqref{eq:DSS_condition_shift}
can be rewritten in terms of~$H_a$. Using
\eqref{eq:DSS_3+1_1}-\eqref{eq:DSS_3+1_2} in
\eqref{eq:dt_alpha_ghg}-\eqref{eq:dt_beta_ghg} gives
\begin{align}
  &\rmd_t \alpha \left(T + \Delta, x^i \right)=\\*
  &- e^{-\Delta} \alpha^2 \left( T, x^i \right)
    \big( n^a \left( T, x^i \right) H_a \left( T + \Delta, x^i \right)
    + K \left(T, x^i \right) \big),\nonumber\\
  &\rmd_t \beta^i \left(T + \Delta, x^j \right)=\\*
  &\alpha^2 \left(T, x^j\right) \gamma^{ik}\left(T, x^j\right)
    \big( H_k \left(T + \Delta, x^j \right) + {}^{(3)}\Gamma_k \left(T, x^j\right) \nonumber\\
  &\quad \quad \quad \quad \quad \quad \quad \quad \quad \quad
    + \partial_k \ln\left(\alpha \left(T, x^j\right) \right) \big).
    \nonumber
\end{align}
So
requiring~\eqref{eq:DSS_condition_lapse}-\eqref{eq:DSS_condition_shift}
is equivalent to requiring
\begin{align}
  \label{eq:DSS_gsf}
  H_a \left(T + \Delta, x^i \right) = H_a \left(T, x^i \right)\, .
\end{align}

The HDWG gauge source function does not satisfy the DSS-compatibility
condition~\eqref{eq:DSS_gsf}, since
\begin{align}
  H^{\text{HDWG}}_a\left(T + \Delta, x^i \right)
  &= -2 \Delta \eta_{L} e^{-\Delta} n_a \left(T, x^i\right)\nonumber\\*
  & \quad+ e^{-\Delta} \eta_{L}
    \ln \left(\frac{\sqrt{\gamma\left(T, x^i\right)}}{\alpha\left(T, x^i\right)}\right)
    n_a\left(T, x^i\right) \nonumber\\*
  & \quad - e^{-\Delta} \eta_{S} \frac{\gamma_{ak}
    \left(T, x^i\right)\beta^k\left(T, x^i\right)}{\alpha\left(T, x^i\right)}
    \nonumber\\*
  &\neq H^{\text{HDWG}}_a \left(T, x^i \right)  \, .
\end{align}

Similarly, neither variant of HDWG, HDWG-$\alpha^2$~\eqref{eq:a2_gsf}
nor HDWG-$\ln(\alpha)$~\eqref{eq:ln_gsf},
satisfies~\eqref{eq:DSS_gsf}. We have instead that
Eq.~\eqref{eq:a2_gsf}
\begin{align}
  \label{eq:a2Ha_noDSS}
  &H^{\text{HDWG-}\alpha^2}_a \left(T + \Delta, x^i \right) = \nonumber\\*
  & \quad \quad-2 \Delta \eta_{L} e^{\Delta} n_a \left(T, x^i\right) \nonumber\\*
  & \quad \quad+ e^{\Delta} \eta_{L}
    \ln\left(\frac{\sqrt{\gamma\left(T, x^i\right)}}{\alpha\left(T, x^i\right)}\right)
    \frac{n_a\left(T, x^i\right)}{\alpha^2\left(T, x^i\right)}\nonumber\\*
  & \quad \quad - \eta_{S} \frac{\gamma_{ak}
    \left(T, x^i\right)\beta^k\left(T, x^i\right)}{\alpha^2\left(T, x^i\right)}
    \nonumber\\*
  &\neq H^{\text{HDWG-}\alpha^2}_a \left( T, x^i \right)
\end{align}
and that Eq.~\eqref{eq:ln_gsf}
\begin{align}
  &H^{\text{HDWG-}\ln(\alpha)}_a \left( T + \Delta, x^i \right)
    =\nonumber\\*
  & \quad e^{-\Delta} n_a\left(T, x^i\right)\Bigg\lbrace\left(-2\Delta + \ln
    \left(\frac{\sqrt{\gamma\left(T, x^i\right)}}{\alpha\left(T, x^i\right)}
    \right)\right)^5 \nonumber \\*
  &\qquad\qquad\qquad\quad
    + \left(-\Delta + \ln \left(\frac{1}{\alpha\left(T, x^i\right)}\right)
    \right)^5\Bigg\rbrace\nonumber\\*
  &\quad - e^{-\Delta}\frac{\gamma_{ak}
    \left(T, x^i\right)\beta^k
    \left(T, x^i\right)}{\alpha\left(T, x^i\right)}\nonumber\\*
  &\qquad\qquad\qquad\quad
    \times\Bigg(-2\Delta + \ln \left(\frac{\sqrt{\gamma\left(T, x^i\right)}}
    {\alpha\left(T, x^i\right)}\right)^4 \Bigg)\nonumber\\*
  &\neq H^{\text{HDWG-}\ln(\alpha)}_a \left(T, x^i \right)\, .
\end{align}
We conclude that none of these popular dynamical gauge conditions that
have recently been used in studies of critical collapse satisfy the
DSS-compatibility condition.

\subsection{DSS-compatible gauge sources}
\label{subsec:our_dss_gsf}
\begin{figure}[t]
  \includegraphics[width=\columnwidth]{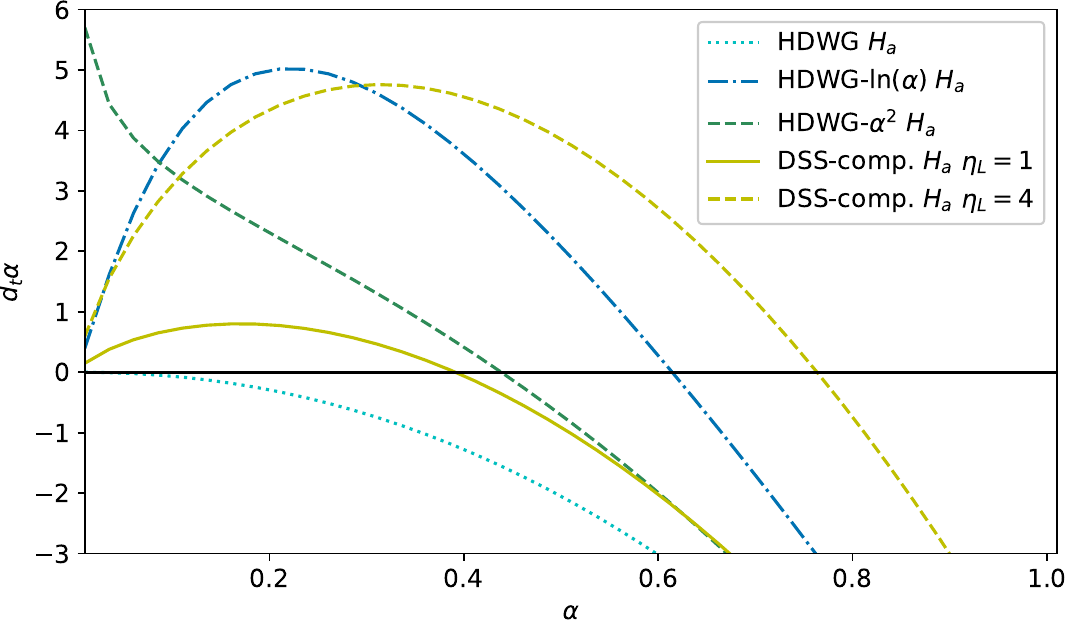}
  \caption{Time evolution equations of the lapse as functions of the
    lapse itself, for the various gauge source functions in
    Eqs.~\eqref{eq:hdwg_dtalpha}, \eqref{eq:ln_dtalpha},
    \eqref{eq:a2_dtalpha} and~\eqref{eq:DSS_dtalpha} with fixed~$K=10$
    and~$\gamma=9$. The coefficients~$\eta_L$ for HDWG and
    HDWG-$\alpha^2$ all taken to be 1 because it is the canonical
    choice and it provides a fair comparison with
    HDWG-$\ln(\alpha)$. We show the slicing condition corresponding to
    the DSS-compatible gauge source function with its~$\eta_{L}$
    being~1 for the comparison, and also 4, a reliable value.}
  \label{fig:slicing_Ha}
\end{figure}

Remaining in the context of GHG from now on, the DSS-compatibility
condition~\eqref{eq:DSS_gsf} can be taken into account to construct
gauge sources designed to satisfy it. As can be seen
in~\eqref{eq:a2Ha_noDSS}, the variant HDWG-$\alpha^2$~$H_a$ only fails
to satisfy~\eqref{eq:DSS_gsf} in the factor proportional to~$\eta_L$,
for~$a=0$. For~$a=i$, the extra~$\alpha^{-1}$, which was initially
added to counteract the collapse of the lapse in the shift evolution
equation of~\eqref{eq:hdwg}, cancels out the factors~$e^{-2\Delta}$
that appear in~$\gamma_{ij}$ after a~$\Delta$ period. However, the
same extra~$\alpha^{-1}$ added to the first term of~\eqref{eq:a2_gsf}
with the same intention but regarding this time the slicing
condition~\eqref{eq:dt_alpha_ghg} fails to satisfy the
DSS-compatibility condition and needs to be adjusted.

As mentioned in Sec.~\ref{subsec:gsf}, the authors of~\cite{LinSzi09}
and~\cite{DepKidSch18} point out the importance of including natural
logarithms of~$\sqrt{\gamma}\alpha^{-1}$ in the slicing condition to
prevent divergences.  Therefore we keep the logarithms, but suggest to
balance out the~$e^{\pm\Delta}$ factors inside the logarithm, in order
to satisfy~\eqref{eq:DSS_gsf}. This leads to
\begin{align}
  \label{eq:DSS_Ha}
  H_a
  &= \eta_L\alpha^{-1} \,\ln(\alpha^{-3}\sqrt{\gamma})n_a
    -\eta_S \,\gamma_{ai}\beta^i \alpha^{-2} \, ,\\
  \label{eq:DSS_dtalpha}
  \rmd_t\alpha
  &= \,\eta_L \alpha \ln(\alpha^{-3}\sqrt{\gamma})
    - \alpha^2 K \, , \\
  \rmd_t\beta^i
  &= -\eta_S \beta^i +\alpha^2 \big({}^{(3)}\Gamma^i
    - \partial^i\ln\alpha\big) \, ,
\end{align}
such that
\begin{align}
  H^{\text{DSS-comp}}_a\left( T +\Delta, x^i\right)
  &= \eta_L \ln\left(\frac{\sqrt{\gamma\left(T, x^i\right)}}
    {\alpha^3\left(T, x^i\right)}\right)
    \frac{n_a\left(T, x^i\right)}{\alpha\left(T, x^i\right)} \nonumber\\
  &\quad -\eta_S \frac{\gamma_{ai}\left(T, x^i\right)\beta^i
    \left(T, x^i\right)}{\alpha^2\left(T, x^i\right)}\nonumber\\
  &=H^{\text{DSS-comp}}_a\left( T , x^i\right)\, .
\end{align}

This is far from a unique choice to fulfill the compatibility
condition. One way to assess whether this choice is reasonable is to
compare it with existing working gauge source
functions. Fig.~\ref{fig:slicing_Ha} shows, for specific values of the
trace of the extrinsic curvature~$K$ and determinant of the spatial
metric~$\gamma$, the profiles of~$\rmd_t\alpha$ as functions of the
lapse~$\alpha$ itself.  In this figure we can see the different ways
in which variations of HDWG deal with a near-vanishing lapse.  For
HDWG one might get the impression that $\rmd_t \alpha$ is always
negative, but in fact it becomes positive, in a small range
near~$\alpha=0$ and the maximal value is also vanishingly small
compared to the other gauges. That means that the HDWG gauge is
counteracting the collapse of the lapse less strongly and also only at
smaller~$\alpha$. The~$\rmd_t \alpha$ corresponding to
HDWG-$\ln(\alpha)$ has a larger maximum and becomes positive at a
larger~$\alpha$, intuitively resulting in pushing~$\alpha$ towards
larger values in an evolution, typically to the region in
which~$\rmd_t \alpha$ changes sign. In the unlikely case of the lapse
somehow dropping to zero exactly it would also stay there. Very close
to~$\alpha=0$ it would be pushed away from zero, but initially only
very slowly.  The~$\rmd_t \alpha$ corresponding to HDWG-$\alpha^2$
instead diverges as $\alpha$ approaches zero, an effect of the lack of
$\alpha^2$ in~\eqref{eq:a2_dtalpha}.  This gauge will therefore push
very small~$\alpha$ more aggressively to the equilibrium point
with~$\rmd_t \alpha = 0$.  Both modifications of HDWG successfully
fight a vanishing lapse at least until it becomes too small to be
dealt with. The DSS-compatible suggestion~\eqref{eq:DSS_dtalpha} keeps
a factor~$\alpha$ in the first term so it results in a very similar
profile to the~$\rmd_t \alpha$ from HDWG-$\ln(\alpha)$. That can be
better seen in Fig.~\ref{fig:slicing_Ha} with the curve corresponding
to the DSS-compatible gauge and~$\eta_L=4$. Coincidentally, that is
also the value we found provided successful simulations.

In section~\ref{sec:cp} we present numerical critical collapse
evolutions using this gauge with the choice of~$\eta_L=4$
and~$\eta_S=6$. Assessing how well these coordinates adapt to DSS can
only be done once the self-similar phase is successfully reached. To
ensure we can stably evolve our simulations up to that point, we next
treat a second major obstruction, namely constraint violations.

\section{The reduction constraint damping scheme}
\label{sec:damping}

Ensuring that the constraints of the system are well satisfied
throughout the evolution guarantees that the numerically evolved data
does truly represent the mathematical and physical problem we set out
to evolve. Since numerical error cannot be completely avoided however,
constraint violations will still occur. What is needed is a strategy
to prevent them from dominating a simulation and potentially even
causing the code to crash.

As suggested in~\cite{BroFriHub98}, some systems of equations admit a
damping scheme that reduces these violations by making the constraint
surface (in solution space) an attractor. This was done explicitly for
the Z4 and GHG systems in~\cite{GunGarCal05, LinSchKid05} by adding to
the evolution equations terms containing multiples of the constraints
and freely specifiable
parameters. In~\eqref{eq:ghg_g}-\eqref{eq:ghg_Phi} these are the terms
with~$\gamma_0, \gamma_1$ and~$\gamma_2$.

As mentioned in section~\ref{sec:intro}, we are primarily concerned
with violations of the reduction constraints, as these are not present
in second order formulations, which yield a finer tuning of the
threshold parameter and so are a point of suspicion. Consequently we
restrain our analysis to a linear model for GHG whose only
non-principal terms are those containing the reduction constraint,
plus linear terms.

\subsection{Mode analysis}
\label{subsec:math}

Taking the first order GHG evolution
equations~\eqref{eq:ghg_g}-\eqref{eq:ghg_Phi} in vacuum, linearizing
about a constraint satisfying background with perturbations that
satisfy the harmonic constraint~\eqref{eq:harmonic_constraint},
working in the constant coefficient approximation and finally assuming
that~$\Pi_{ab}=\Phi_{iab}=0$ in the background gives
\begin{align}
  \label{eq:reduced_ghg_g}
  \partial_t \tensor{g}{_a_b}
  &=\beta^i \partial_i \tensor{g}{_a_b}
    - \alpha \tensor{\Pi}{_a_b}
    + \gamma_1 \beta^i \tensor{C}{_i_a_b}\, , \\
  \label{eq:reduced_ghg_Pi}
  \partial_t \tensor{\Pi}{_a_b}
  &=\beta^i \partial_i \tensor{\Pi}{_a_b}
    - \alpha \gamma^{ij} \partial_i \tensor{\Phi}{_j_a_b}
    + \gamma_1 \gamma_2 \beta^i \tensor{C}{_i_a_b}\, ,\\
 \label{eq:reduced_ghg_Phi}
  \partial_t \tensor{\Phi}{_i_a_b}
  &=\beta^j \partial_j \tensor{\Phi}{_i_a_b}
    - \alpha \partial_i \tensor{\Pi}{_a_b}
    + \gamma_2 \alpha \tensor{C}{_i_a_b} \,,
\end{align}
where, here and for the rest of this section, we overload the notation
so that~$g_{ab}$, and so forth, stand for the metric perturbation such
that~\eqref{eq:reduced_ghg_g}-\eqref{eq:reduced_ghg_Phi} holds up to
first order in the perturbation.  The remaining
coefficients~$\alpha,\beta^i$ and~$\gamma^{ij}$ are the constant
lapse, shift and spatial metric in the background.  These simplifying
assumptions suppress any coupling between different components, making
the following analysis algebraically tractable. Here the analysis
deviates from that done in~\cite{GunGarCal05} as the lapse and the
shift are not taken to be $1$ and $0$ respectively but instead to be
constant in the background. It differs from the analysis provided
in~\cite{LinSchKid05} as we here take into account non-principal
terms. Although it would be interesting to do so, dropping any of the
remaining assumptions would make the computation very substantially
more complicated and, since we are only trying to motivate a simple
change to our constraint damping scheme, does not seem worthwhile
here.

A mode analysis of the evolution system can tell us the rate at which
the constraint violations grow in this approximation.
Writing~$\mathbf{u} = (g_{ab}, \Pi_{ab}, \Phi_{iab})^T$, the system of
equations~\eqref{eq:reduced_ghg_g}-\eqref{eq:reduced_ghg_Phi} can be
written as
\begin{align}
  \label{eq:matrix_rep}
  \partial_t \mathbf{u} =
  \mathbf{A}^{k} \partial_k\mathbf{u} + \mathbf{B} \mathbf{u}\,.
\end{align}
Here~$\mathbf{A}^{k}$ correspond to the principal part matrices of
GHG, whereas~$\mathbf{B}\mathbf{u}$ corresponds to the subset of
non-principal terms that survive linearization and our simplifying
assumptions. Observe that the~$4$d indices~$ab$ can be omitted for the
analysis.

Moving to the frequency domain, we make a mode
Ansatz~$\tilde{\mathbf{u}}= \tilde{\mathbf{u}}_0 \, e^{st+i\omega_k
  x^{k}}$ where~$\tilde{\mathbf{u}}$ is the Laplace-Fourier transform
of~$\mathbf{u}$, $\omega_k =|\omega|\hat{\omega}_k$ is an
arbitrary~$3$d-vector in the frequency domain, and we
write~$|\omega| = \sqrt{\gamma^{ij}\omega_i \omega_j }$. The
system~\eqref{eq:matrix_rep} then takes the form of the following
eigenvalue problem
\begin{align}
  \label{eq:eigenpbm}
  s \tilde{\mathbf{u}} = \mathbf{M} \tilde{\mathbf{u}}, 
\end{align}
where~$\mathbf{M} = i \mathbf{A}^k\omega_k+\mathbf{B}$.

Using the unit frequency vector~$\hat{\omega}_i$
($\hat{\omega}_i\hat{\omega}^i = 1$) and its orthogonal~$2$d
projection operator~$q_{ij}$ ($\hat{\omega}^i q_{ij}=0$) we can
further simplify the system with a~$2+1$ decomposition
\begin{align}
  \gamma_{ij} = \hat{\omega}_i\hat{\omega}_j + q_{ij},
  \quad \beta^{i} = \hat{\omega}^{i} \beta^{\hat{\omega}}
  + q^{i}{}_A \beta^{A},
\end{align}
where capital Latin letters~$A,B$ denote a projection by~$q_{ij}$,
running over the~$2$d plane orthogonal to~$\hat{\omega}_k$. Our~$2+1$
notation denotes components in the direction of~$\hat{\omega}_i$
as~$\beta^{\hat{\omega}} = \hat{\omega}_{k} \beta^k$ and~$2$d
projected ones as~$\beta^{A} = q^{A}{}_k \beta^k$. Any other vector or
covector is decomposed using an analogous notation. In this notation
the mode variables in the eigenvalue problem~\eqref{eq:eigenpbm}
become~$\tilde{\mathbf{u}}=(\tilde{g}, \tilde{\Pi},
\tilde{\Phi}_{\hat{\omega}}, \tilde{\Phi}_{A})^T$.  The
matrix~$\mathbf{M}$ becomes
\begin{align}
  \label{eq:M}
  \left(
  \begin{array}{cccc}
    i (1 + \gamma_1) |\omega|\beta^{\hat{\omega}} & -\alpha & -\gamma_1 \beta^{\hat{\omega}}  & - \gamma_1 \beta^{A}   \\
    i \gamma_1\gamma_2 |\omega| \beta^{\hat{\omega}} & i|\omega|\beta^{\hat{\omega}} & -i\alpha |\omega| -\gamma_1\gamma_2 \beta^{\hat{\omega}} & -\gamma_1\gamma_2 \beta^{A} \\
    i\alpha \gamma_2 |\omega| & -i \alpha |\omega| & i|\omega| \beta^{\hat{\omega}} - \alpha \gamma_2 & 0   \\
    0 & 0 & 0 & i| \omega| \beta^{\hat{\omega}} - \alpha \gamma_2   
  \end{array}
                \right)\, .
\end{align}
The eigenvalues of~$\mathbf{M}$ are
\begin{align}
  \label{eq:speeds}
  s_{1} &= i|\omega| (\beta^{\hat{\omega}}-\alpha),
          \quad s_{2} =i|\omega| (\beta^{\hat{\omega}}+\alpha),
          \nonumber\\*
  s_{3,4} &=i|\omega|\beta^{\hat{\omega}} -\alpha \gamma_2,
            \quad s_{5} =i|\omega|\beta^{\hat{\omega}}(1+\gamma_1)
            -\alpha \gamma_2.
\end{align}
The imaginary parts of these eigenvalues represent the speeds of the
system's propagation modes and the real parts capture exponential
decay or growth.  The expected speed of light is captured by~$s_1$
and~$s_2$ for $|\omega| = 1$. More interesting to our analysis is the
role of the parameters~$\gamma_1$ and~$\gamma_2$ in the evolution of
the constraints.

We construct the propagation
modes~$v = \mathbf{l}\cdot\tilde{\mathbf{u}}$ from the left
eigenvectors~$\mathbf{l}$ of~$\mathbf{M}$ (see
Appendix~\ref{app:eigen}).  In the generic
case~$\beta^{\hat{\omega}}\ne0$, the propagation modes, whose speeds
contain damping terms, are
\begin{align}
  v_{s_{3,4}}
  &= \tilde{\Phi}_{Aab} = q_{A}{}^i \tilde{\Phi}_{iab} = - \tilde{C}_{Aab} \, ,\\
  v_{s_5}
  &=\beta^{A}\tilde{\Phi}_{Aab} +\beta^{\hat{\omega}}
    (\tilde{\Phi}_{\hat{\omega}ab} - i|\omega|\tilde{g}_{ab}) =
    - \beta^{i} \tilde{C}_{iab} \,,
\end{align}
where for the last equalities we exploited the orthogonality
between~$q_{A}{}^i$ and $\hat \omega_i$
in~$|\omega|\hat \omega_i \tilde g_{ab}$.  This tells us that some
modes of the reduction constraints propagate as~$v_{s_{3,4}}$ with
speed~$s_{3,4}$, whereas some propagate as~$v_{s_{5}}$ with
speed~$s_{5}$. We can then confirm by direct computation
that~$\tilde{C}_{iab}$ is damped directly by a
factor~$-\alpha\gamma_2$. Similar results hold in the special
cases~$\beta^{\hat{\omega}}=0$ and~$\gamma_1=0$.

Despite the simplistic nature of this treatment it helps us to
understand that there is a clear relation between moments where the
lapse collapses and the growth of the constraints, as we explain in
the next section.

\subsection{Adjusted damping}
\label{subsec:adjusted_damp}
\begin{figure}[t]
  \includegraphics[width=\columnwidth]{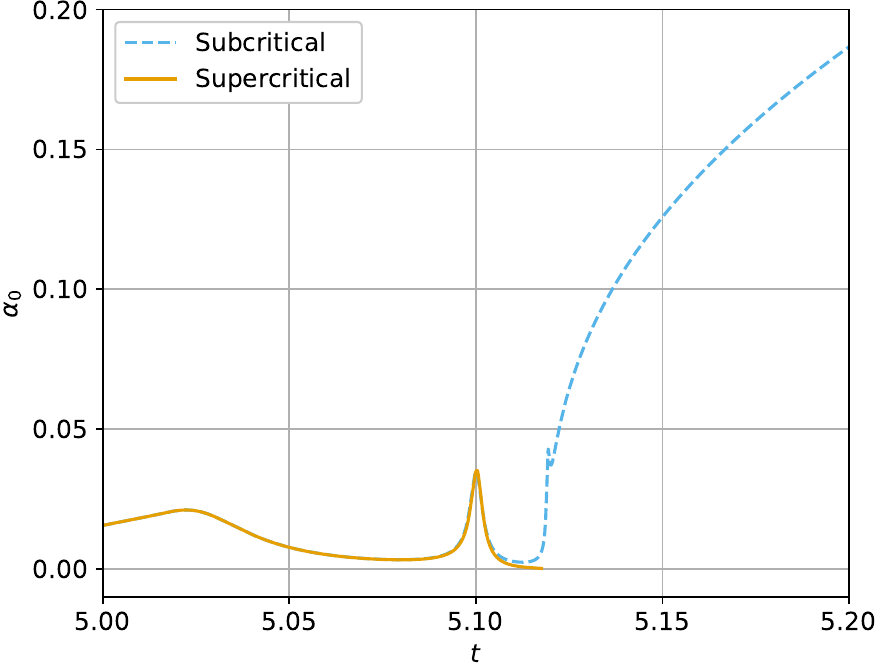}
  \caption{The value of the lapse at the origin is plotted as a
    function of time in simulations of moment of time symmetry scalar
    field initial data. The orange line corresponds to a supercritical
    simulation that stopped and an apparent horizon was found at the
    time we observe the lapse reaches~$0$. The blue dashed line
    corresponds to a subcritical simulation where, despite a large
    peak of curvature exactly when the lapse approaches~$0$, the
    fields manage to eventually disperse.}
  \label{fig:collapse_lapse}
\end{figure}
\paragraph{Collapsing lapse}
\label{subsubsec:collapse_lapse}

The mode analysis shows that violations of the reduction constraints,
$C_{iab}$, are damped at a rate~$-\alpha\gamma_2$. In the limit where
the lapse tends to zero, constraint violations are no longer damped
and can start to grow due to numerical round-off. It is an empirical
fact that, with many popular gauge conditions, the lapse~$\alpha$
collapses to zero in the presence of an apparent horizon. Although
this is a coordinate dependent feature, it is so consistent that a
collapsing lapse is often used to detect the presence of a black
hole. To the best of our knowledge, there is no mathematical proof in
which this is demonstrated for GHG, but we do observe such behavior
numerically. Intuitively, this means that spatial slices stretch and
wrap around a singularity, causing a rapid growth of the radial metric
components whilst proper time gradually freezes inside the horizon,
but advances on the outside. This time freezing effect near a
singularity can even be desirable for certain simulation purposes, for
instance to steer clear of the singularity.

Although pure harmonic slicing is only ``marginally singularity
avoiding'' in the terminology of Alcubierre~\cite{Alc02}, our
experience with GHG in \texttt{bamps} is that every time we have
confirmed the detection of an apparent horizon with either harmonic or
one of the gauge source functions discussed above, the lapse had
collapsed almost to zero, as Fig.~\ref{fig:collapse_lapse}
shows. Interestingly, we observe the same tendency when the curvature
is extreme, even in the absence of an apparent horizon. For instance
in evolutions of subcritical data close to the threshold of
gravitational collapse, we observe a nearly vanishing lapse at the
time where the curvature invariant peaks. A crucial example of this
are our studies of critical phenomena in gravitational collapse.

\begin{figure*}[t]
	\includegraphics[width=1 \textwidth]{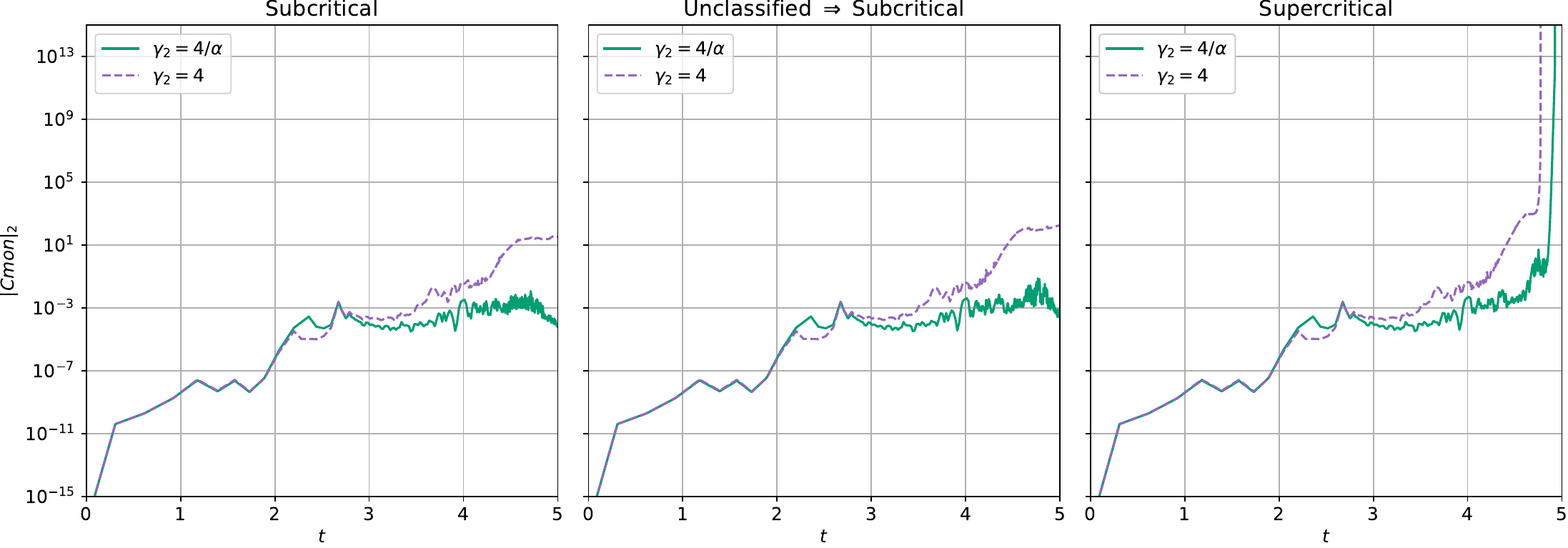}
	\caption{The~$L_2$ norm of the sum of all constraints (the
          ``constraint monitor") with and without the adjusted damping
          parameter; in the left for a subcritical amplitude; in the
          middle for an amplitude that was not classified without the
          adjusted damping of Eq.~\eqref{eq:damp_hack}, but could be
          classified as subcritical with the adjusted damping; and a
          supercritical amplitude. The data used in these plots are
          for configurations~\ref{id:MTS_g4oalpha_err_ln}
          and~\ref{id:MTS_g4_err_ln} in table~\ref{table:all_in_one}.
        }
	\label{fig:Cmon}
\end{figure*}
\paragraph{Adjusted damping parameters}
\label{subsubsec:collapse_proof_damp}

Given that the rate of damping of~$C_{iab}$ is suppressed as the lapse
collapses, and that in critical phenomena simulations, close to the
threshold, the lapse consistently collapses, we hypothesize that this
vanishing lapse results in the large constraint violations we observe
close to the threshold. This may even result in code failures before
field dispersion or trapped surface formation, thereby obstructing the
threshold parameter search.  This could account for at least a subset
of the shortcomings of first order GHG in the tuning of the threshold
parameter, as there is no reduction constraint to deal with in second
order systems. To test our hypothesis we adjusted the
parameter~$\gamma_2$ in the GHG
equations~\eqref{eq:ghg_g}-\eqref{eq:ghg_Phi} such that the damping
terms in $s_{3,4}$ and $s_{5}$ decouple from the lapse
\begin{align}
  \label{eq:damp_hack}
  \gamma_2 = c_{\gamma_2} \, \rightarrow \, \gamma_2
  = \frac{c_{\gamma_{2}}}{\alpha} \, ,
\end{align} 
with typically $c_{\gamma_2}=2$ or~$4$.

In this way, the damping of the reduction constraint violation speeds
can resist a collapsing lapse
\begin{align}
  s_{3,4} &{}= i|\omega|\beta^{\hat{\omega}}
            - c_{\gamma_{2}} \, , \\
  s_{5}   &{}= i|\omega|\beta^{\hat{\omega}}(1+\gamma_1)
          - c_{\gamma_{2}}.
\end{align} 
Essentially with this change the damping timescale is given in
terms of coordinate, rather than proper time.

\paragraph{Damped Constraints}
\label{subsubsec:damp_works}

As shown in Fig.~\ref{fig:Cmon}, using the adjusted damping reduces
the constraint violation of both subcritical and supercritical
evolutions as desired. Crucially, we can see that it holds the
constraints low enough for a previously unclassified spacetime to be
classified as subcritical, enabling the continuation of the parameter
search. In section~\ref{sec:cp} we now demonstrate that this
improvement helps in the observation of critical phenomena using first
order GHG.

\section{Critical collapse results}
\label{sec:cp}

We now present critical collapse results for two one-parameter
families of spherical initial data for the massless scalar field. We
treat these spacetimes as a (very challenging) testbed for the code,
and as such assume some familiarity with the standard notions of
critical collapse. We refer the reader to~\cite{GunGar07} for a
comprehensive overview. The first family is the moment of time
symmetry data, Eqs.~\eqref{eq:initial_phi_mts}
and~\eqref{eq:initial_pi_mts}, the second incoming initial data ,
Eqs.~\eqref{eq:initial_phi_inc} and~\eqref{eq:initial_pi_inc}.  All
bisection searches to the critical amplitude follow the same
procedure. We classify data as subcritical if it evolves until the
fields are fully dispersed. Although our code supports a robust method
for spherical black hole excision~\cite{BhaHilRaj21}, we are presently
only interested in knowing whether or not an apparent horizon forms,
so do not treat the subsequent black hole evolution carefully. The
code therefore eventually fails for all supercritical data. In
spherical symmetry, the presence of apparent horizons can be
determined by zero-crossings of the expansion, which is algebraic in
our evolved variables~\cite{Alc08}. We only classify failed evolutions
as supercritical if they display such a negative expansion. Starting
with a sufficiently wide initial window, we proceed to bisect,
evolving data whose amplitude is in the middle of the regime. If the
evolution is subcritical we update the lower bound and it if its
supercritical we update the upper bound. We proceed with the bisection
until we reach data that cannot be classified. In all plots the slow
time~$T_p$, defined in~\eqref{eq:slow_time_prop}, is computed from the
proper time at the origin as
\begin{align}
  \label{eq:tau}
  \tau(t') = \int_0^{t'} \alpha(t, 0) dt 
\end{align}
and the accumulation time~$\tau_{*}$ following Eq.~(23)
in~\cite{Bau18}. We computed the period~$\Delta$ using Eq.~(24)
in~\cite{Bau18}, Similarly, we computed the proper length~$x_p$ from
the $x$-coordinate as
\begin{align}
	\label{eq:xp}
	x_p(x') = \int_0^{x'} \sqrt{g_{xx}(t,x)} dx \,  .
\end{align}

\subsection{Constraint violations}
\begin{table*}
\begin{ruledtabular} 
\begin{tabular} { l|llllll }
  & Identifier & $\gamma_2$ & \shortstack{indicator for\\h-refinement} & $H_a$ &   $ A_{\text{sub}} $ & $ A_{\text{sup}}$ \\
  \hline
  \multirow{6}{45pt}{Moment of time symmetry data} 
  & \ID{1}{id:MTS_g4oalpha_smoo_HDWG} 
  & $4/\alpha$ & smoothness & HDWG & $ 0.15(0000000000000) $ & $ 0.15(6250000000000) $\\ 
  & \ID{2}{id:MTS_g4oalpha_smoo_alpha2} 
  & $4/\alpha$ & smoothness & HDWG-$\alpha^2$ & $0.151675332247(121) $ & $ 0.151675332247(212) $\\
  & \ID{3}{id:MTS_g4_err_ln} 
  & $4$        & error & HDWG-$\ln(\alpha)$ & $ 0.151673(126220703) $ & $ 0.151673(889160156) $  \\
  & \ID{4}{id:MTS_g4oalpha_err_ln} 
  & $4/\alpha$ & error & HDWG-$\ln(\alpha)$ & $ 0.15167533190(5423) $ & $ 0.15167533190(6879) $  \\ 
  & \ID{5}{id:MTS_g4oalpha_smoo_ln} 
  & $4/\alpha$ & smoothness & HDWG-$\ln(\alpha)$ &  $0.151675332(244484) $ & $ 0.151675332(337617) $ \\
  & \ID{6}{id:MTS_g4oalpha_smoo_DSS} & $4/\alpha$ & smoothness & DSS-comp. &  $ 0.1516753322(44484) $ & $ 0.1516753322(91050) $\\
 \hline
 \multirow{6}{45pt}{Incoming data} 
  & \ID{7}{id:Inc_g2oalpha_smoo_HDWG} 
  & $2/\alpha$ & smoothness & HDWG & $ 0.1(00000000000000) $ & $ 0.1(18750000000000) $  \\
  & \ID{8}{id:Inc_g2oalpha_smoo_alpha2} 
  & $2/\alpha$ & smoothness & HDWG-$\alpha^2$ & $ 0.10893328143(4289) $ & $ 0.10893328143(6472) $  \\
  & \ID{9}{id:Inc_g2_smoo_ln} 
  & $2$        & smoothness & HDWG-$\ln(\alpha)$ & $ 0.1089332(75938033) $ & $ 0.1089332(84878729) $  \\
  & \ID{10}{id:Inc_g2oalpha_smoo_ln} 
  & $2/\alpha$ & smoothness & HDWG-$\ln(\alpha)$ &  $0.108933281(395000) $ & $ 0.108933281(403731) $  \\
  & \ID{11}{id:Inc_g2oalpha_smoo_DSS} 
  & $2/\alpha$ & smoothness & DSS-comp. &  $ 0.1089332815(08505) $ & $0.1089332815(25968) $  \\
\end{tabular}
\end{ruledtabular} 
\caption{Bisection intervals for various configurations.  The critical
  amplitude~$A_{*}$ is located in the
  interval~$[A_{\text{sub}}, A_{\text{sup}}]$. The identifier serves
  as reference for the configurations in the text. $\gamma_2$ is the
  constraint damping factor for the reduction constraints, see
  Eqs.~\eqref{eq:ghg_g}-\eqref{eq:ghg_Phi}. $H_a$ indicates the choice
  of gauge source function, which are introduced in
  Sec.~\ref{subsec:gsf} and Sec.~\ref{subsec:our_dss_gsf}.  }
\label{table:all_in_one}
\end{table*}
\begin{figure*}[t]
  \includegraphics[width=\columnwidth]{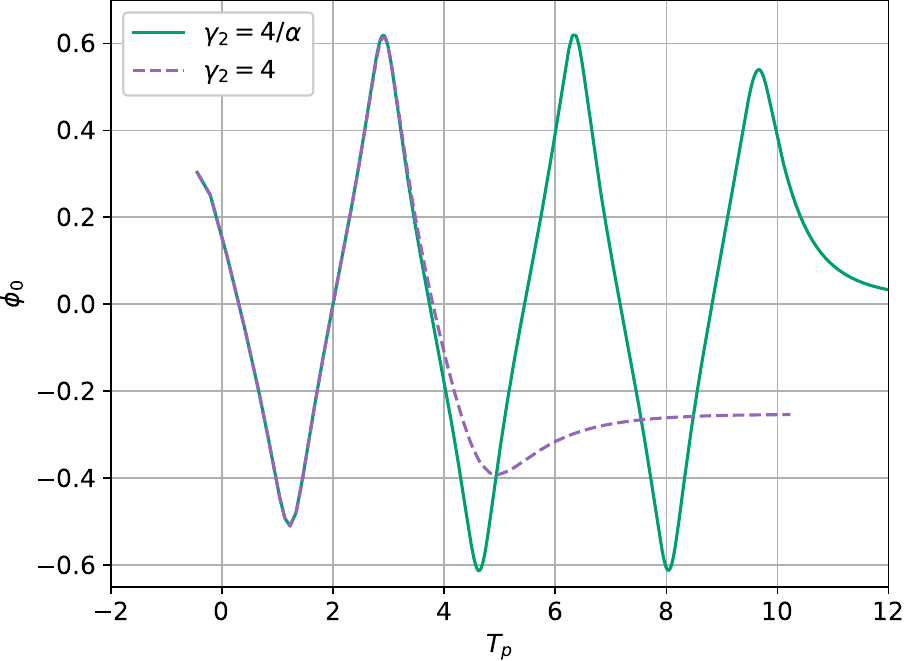}
  \includegraphics[width=\columnwidth]{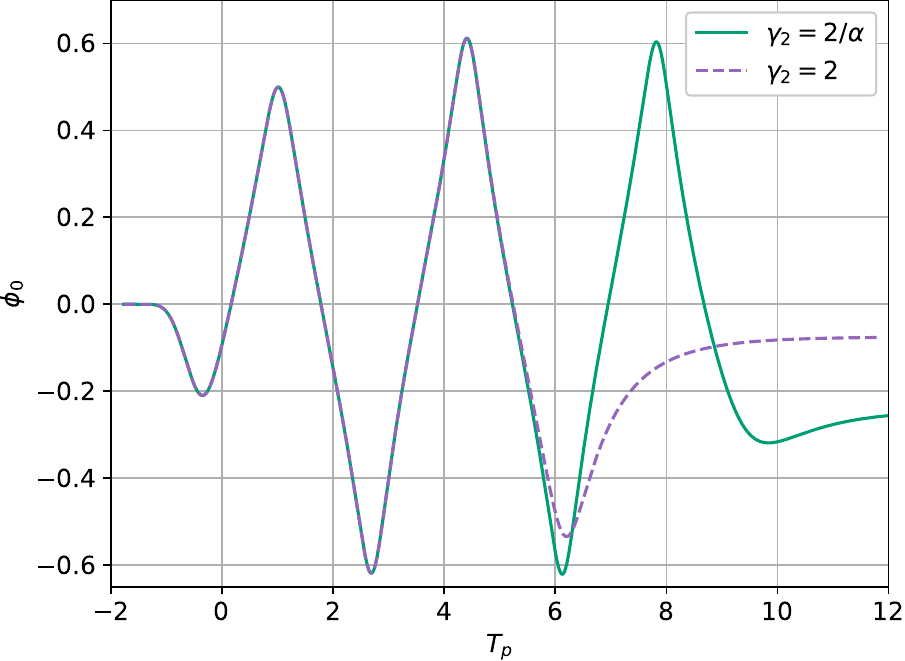}
  \caption{On the left we show echoes of the highest subcritical
    amplitudes of two moment of time symmetry data bisections: one
    with the original damping parameter
    (configuration~\ref{id:MTS_g4_err_ln}) and one with the adjusted
    damping parameter of Eq.~\eqref{eq:damp_hack}
    (configuration~\ref{id:MTS_g4oalpha_err_ln}).  The lack of
    reliable tuning beyond~6 digits is an obstacle to see critical
    phenomena with the original damping scheme. On the right we show
    the same for the incoming initial data family
    (configuration~\ref{id:Inc_g2oalpha_smoo_ln}
    and~\ref{id:Inc_g2_smoo_ln}). \label{fig:damp_no_damp_echoes} }
\end{figure*}

Critical phenomena of a massless scalar field minimally coupled to GR
in spherical symmetry have been very well studied and reproduced. It
was therefore surprising to us that our initial automated bisections
of moment of time symmetry data with the standard first order GHG
constraint damping setup with HDWG-$\ln(\alpha)$
$H_a$~\eqref{eq:ln_gsf}, apparently tuning up to 15 digits, displayed
at most one and a half echoes. Closely looking at the data, large
constraint violations were causing crashes and noisy data caused
incidental negative values of the expansion. The 15 digit tuning was
not trustworthy, and the solution space had drifted towards the
subcritical side by classifying as supercritical what was simply noisy
data without horizons. The last reasonable estimation would only be
after a 6 digit tuning, already giving indeed the same echo and a half
we observed with the initially wrongly-tuned 15 digits.

We obtained comparably poor results with the incoming data, only
reliably tuning~7 digits. Those were our best results after increasing
the damping parameters and forcing more refinement (to no avail). For
this reason the tables and figures mention different refinement
indicators and values of~$\gamma_2$ for the two different families of
data. With our initial gauge source function HDWG-$\alpha^2$,
Eq.~\eqref{eq:a2_gsf}, we obtained similarly large constraint
violations and unreliable results far from the 3 echoes
that~\cite{Bau18} found.

By instead choosing our adjusted damping parameter of
Eq.~\eqref{eq:damp_hack}, immune to the collapsing lapse that occurs
as curvature approaches the supercritical side, we were able to
control those constraint violations. As Fig.~\ref{fig:Cmon} shows,
close to the time of highest curvature ($t\approx 5$ for those
simulations), the constraints are damped much more effectively when
using the adjusted damping factor.

This improvement of the constraints near criticality has allowed us to
confidently classify data on both sides of the threshold. From the
subcritical side, it avoids crashes caused by large constraint
violations, allowing the simulation to last until the fields disperse,
as the central plot in Fig.~\ref{fig:Cmon} shows. From the
supercritical side, we observe clear and smooth indications of
apparent horizons in the evolution of the expansion.

It should be stressed that the adjusted parameter does not provide an
improvement of the constraints throughout the entire evolution, as can
be seen at~$t\approx 2.5$ in Fig.~\ref{fig:Cmon} or at later
times. Instead, it makes a clear improvement in the key case where the
lapse collapses and the spacetime starts displaying critical
phenomena, at~$t\approx 5$ in this case.

Table~\ref{table:all_in_one} provides an overview over all
configurations discussed in the text. The table shows that the
adjusted constraint damping parameter improved the estimation of the
critical amplitude from 7 to 9 digits for the incoming data family
(compare configurations~\ref{id:Inc_g2_smoo_ln}
and~\ref{id:Inc_g2oalpha_smoo_ln}) and from 6 to 11 digits for the
moment of time symmetry family (compare
configurations~\ref{id:MTS_g4_err_ln}
and~\ref{id:MTS_g4oalpha_err_ln}).  As seen
in~Fig.~\ref{fig:damp_no_damp_echoes}, this has allowed us to observe
up to 3 echoes.

\begin{figure}[t]
  \includegraphics[width=\columnwidth]{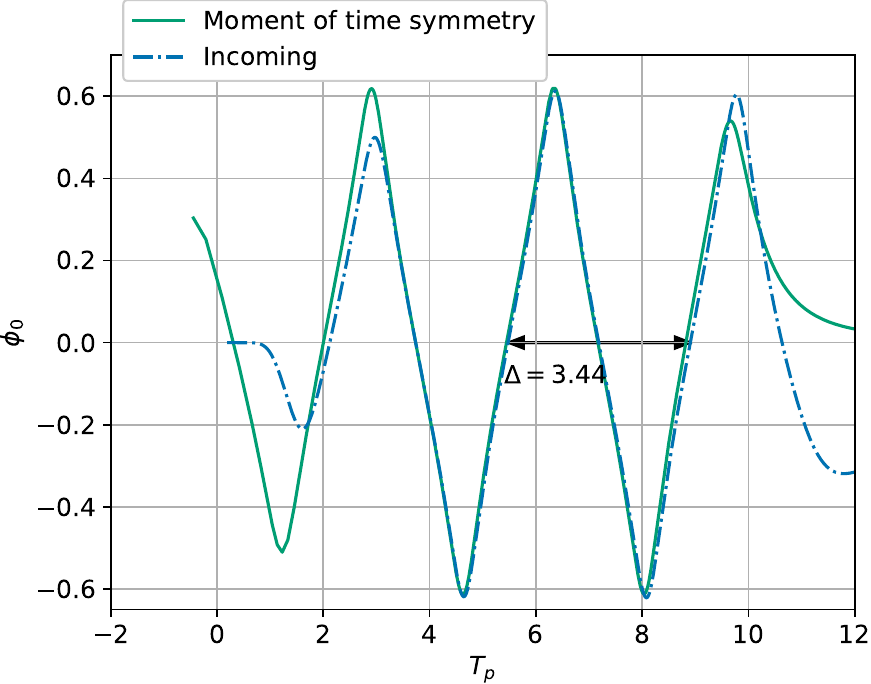}
  \caption{Echoes of the highest subcritical amplitudes of incoming
    initial data (configuration~\ref{id:Inc_g2oalpha_smoo_ln}) and
    moment of time symmetry data
    (configuration~\ref{id:MTS_g4oalpha_err_ln}) and the adjusted
    damping of Eq.~\eqref{eq:damp_hack}. The incoming data has been
    shifted to the right to enable a direct comparison with the moment
    of time symmetry data.\label{fig:data_echoes_arrow} }
\end{figure}

Fig.~\ref{fig:data_echoes_arrow} shows 3 echoes for both the moment of
time symmetry data and the incoming data. The former becomes
approximately self-similar slightly earlier, as its first echo is
stronger than the incoming data's first echo, whilst its last one is
incomplete compared to the incoming data. In this plot the incoming
data has been shifted to the right to verify that, indeed, different
initial data displays the same echoing near criticality with a period
of~$\Delta\simeq3.44$. We conclude that the adjusted damping was
essential to see critical phenomena with GHG.

\subsection{Coordinate behavior}
\label{subsec:coord}
\begin{figure}[t]
  \includegraphics[width=\columnwidth]{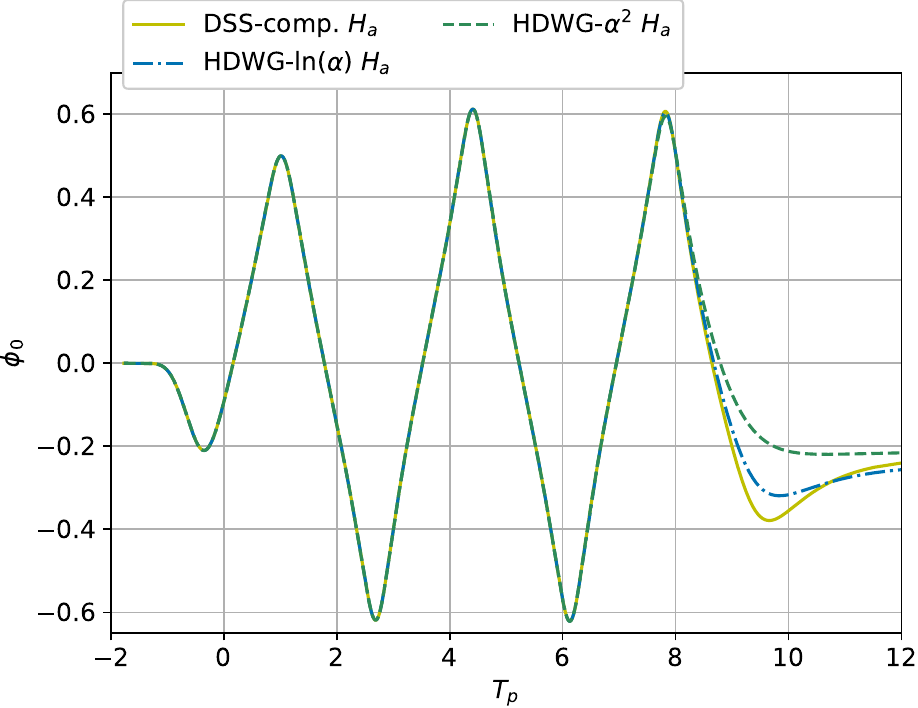}
  \includegraphics[width=\columnwidth]{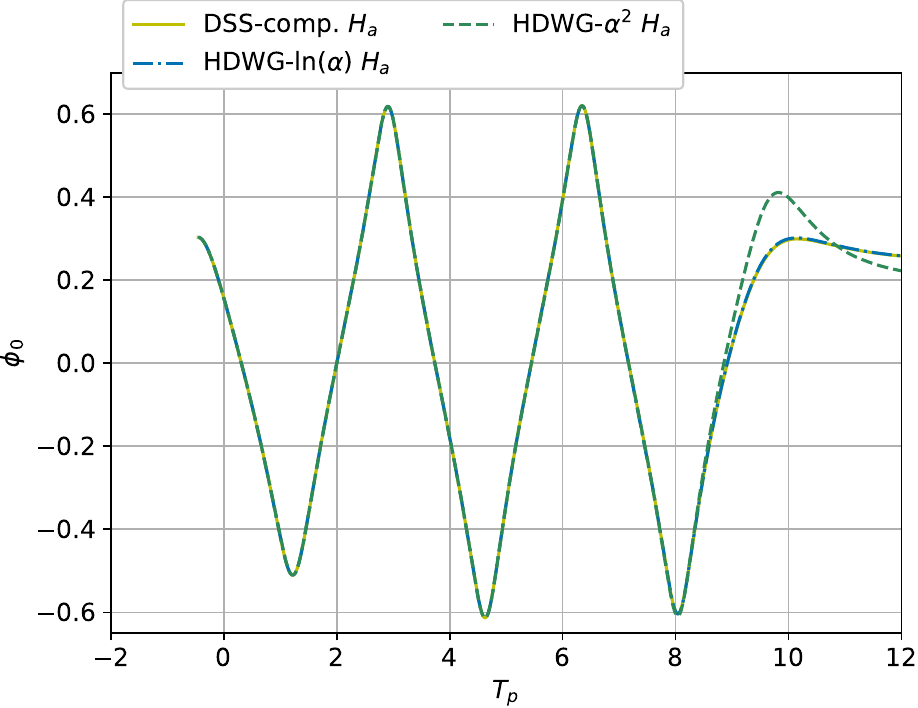}
  \caption{Echoes of the highest subcritical amplitudes of incoming
    data bisections performed with the adjusted damping of
    Eq.~\eqref{eq:damp_hack} and all the mentioned gauge source
    functions. The top panel shows incoming data
    (configurations~\ref{id:Inc_g2oalpha_smoo_DSS},
    \ref{id:Inc_g2oalpha_smoo_ln}
    and~\ref{id:Inc_g2oalpha_smoo_alpha2}) while the lower panel shows
    moment of time symmetry data
    (configurations~\ref{id:MTS_g4oalpha_smoo_DSS},
    \ref{id:MTS_g4oalpha_smoo_ln}
    and~\ref{id:MTS_g4oalpha_smoo_alpha2}). The DSS-compatible gauge
    source gives results comparable to the ones obtained with
    HDWG-variants, being therefore also suited for critical collapse
    simulations.
    \label{fig:gsf_echoes}    
  }
\end{figure}

Once our bisections reach spacetimes that are close enough to
criticality, we can evolve those up to the self-similar phase and
evaluate the utility of the DSS-compatible gauge source function.

We perform bisections of the same data using different gauge source
functions in the evolutions, namely HDWG~\eqref{eq:hdwg},
HDWG-$\alpha^2$~\eqref{eq:a2_gsf},
HDWG-$\ln(\alpha)$~\eqref{eq:ln_gsf} and finally our initial
suggestion of a DSS-compatible gauge source
function~\eqref{eq:DSS_Ha}.

For this comparison we employ the same refinement strategy for both
sets of data using the smoothness indicator for h-refinement
(see~\cite{RenCorHil23} for details). This provides low constraint
violations with all gauge combinations but does actually display a
lower peak in the third echo for the moment of time symmetry data
(compare the lower panel of Fig.~\ref{fig:gsf_echoes} with the
straight line in Fig.~\ref{fig:data_echoes_arrow}). The self-similar
phase is well approached for all combinations once the adjusted
damping parameter is used. As can be seen by comparing
configuration~\ref{id:MTS_g4oalpha_err_ln}
and~\ref{id:MTS_g4oalpha_smoo_ln} in table~\ref{table:all_in_one} the
values of the critical regime do vary within a family with different
h-refinement strategies after the 8th digit.

Table~\ref{table:all_in_one} shows the intervals that contain the
critical amplitude~$A$ from the bisections. The values
of~$A_{\text{sub}}$ correspond to the curves in
Fig.~\ref{fig:gsf_echoes}. These runs have been checked and are not
always the raw output of the automated bisections. In particular, like
in the case discussed in the previous subsection, the automated
bisection of the moment of time symmetry data evolved with
HDWG-$\alpha^2$ gauge source function
(configuration~\ref{id:MTS_g4oalpha_smoo_alpha2}) gave a misleading~14
digit tuning that actually showed no improvement in the echoing and
large constraint violations after the 12th digit. We therefore trust
the bisection up to the 12th digit only.

It is expected that the use of different coordinates not only affect
how close we can evolve to the threshold but it also can drift the
solution space slightly because of different numerical error. For
example in table~\ref{table:all_in_one} the critical intervals
obtained with HDWG-$\alpha^2$
(configuration~\ref{id:Inc_g2oalpha_smoo_alpha2}) and the
DSS-compatible gauge (configuration~\ref{id:Inc_g2oalpha_smoo_ln})
differ in their 10th digit but none is in principle more real. The
true value could perhaps be more accurately obtained with careful
convergence testing, but since round-off error may already contribute
with this level of tuning, and we are interested here primarily in
capturing the correct phenomenology, we make no attempt to do so.

We can use again Fig.~\ref{fig:gsf_echoes} to assess from which point
this difference in tuning stops being meaningful with respect to the
observed critical phenomena. For both sets of data, the low level of
tuning obtained with vanilla HDWG (of only 1 or 2 digits) had a clear
impact on our capacity to observe critical phenomena as it does not
give any echoes for either data type. The difference in tuning
obtained with vanilla HDWG compared to the other gauge source
functions is therefore a severe shortcoming. Moving on to the other
three gauge source functions, the lower panel of
Fig.~\ref{fig:gsf_echoes} displaying moment of time symmetry data
shows that the better tuning obtained with the HDWG-$\alpha^2$ variant
does indeed give the highest last echo. There both the data evolved
with HDWG-$\ln(\alpha)$ and the DSS-compatible version behave in
exactly the same way, so the difference in the tuning they provide,
although not unphysical, is not significant. In the upper panel
displaying incoming data, the best tuned data, also obtained with
HDWG-$\alpha^2$ $H_a$, this time does not provide the best last
echo. It comes third after the one obtained with HDWG-$\ln(\alpha)$
$H_a$ and the DSS-compatible version. The improved tuning obtained
with HDWG-$\alpha^2$ is therefore not as meaningful as that obtained
with the DSS-compatible gauge for this family of data. Again the
constraint violations are all comparable so all critical intervals are
valid, but it does seem that gauge differences affect how easily,
after how many digits, we can evolve near criticality.

A sharp profile in the shift was particularly prominent when evolving
vanilla HDWG with both sets of data. In order to examine coordinate
behaviour more closely, we compare the last common amplitude run in
bisections of all three different gauge source functions that approach
criticality. For the incoming data this corresponds to evolutions
of~$A=0.108933281(456118)$. The evolution of this data with the
DSS-compatible gauge shows a clear uninterrupted dispersion of the
fields allowing to classify the spacetime as subcritical. In contrast,
both adjustments to HDWG crash, classifying it as supercritical. A
common feature to both of the HDWG-variant evolutions is the presence
of what resembles a step function in the profile of the shift vector
in space. The constraint violations are comparable enough to trust
both classifications but we see that these sharp features in the shift
seem to appear later in the bisection for this family of data with the
DSS-compatible guage source function. This advantge does not seem to
be the case for the last common amplitude in the moment of time
symmetry data, $A=151675332(291050)$ which shows very similar profiles
of the shift, none as smooth as would be desirable, for bisections
with all three gauge source functions. This spacetime was classified
as subcritical with HDWG-$\ln(\alpha)$ $H_a$ whereas a horizon was
found with the other two choices.

To summarize, the two HDWG-variants considered and the DSS-compatible
gauge agree very well to to 9 digits for both sets of data with
comparable critical results and constraint violations. With this level
of tuning our results will be affected by round-off, so to tune
further we would need to work with higher precision arithmetic. More
work will be needed to completely avoid the gauge features that affect
the bisections near criticality, but we nevertheless conclude that the
DSS-compatible gauge source function maintains well behaved
coordinates far into the critical regime.

\subsection{DSS and gauge sources }
\label{subsec:DSS_results}
\begin{figure*}[t]
  \includegraphics[width=\columnwidth]{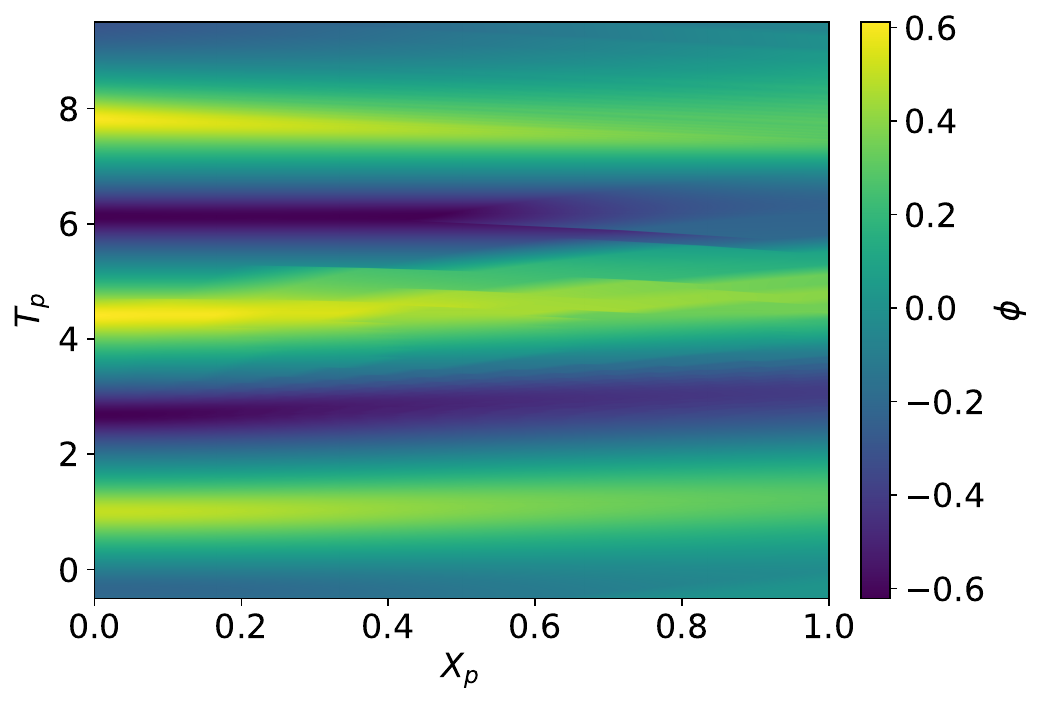}
  \includegraphics[width=\columnwidth]{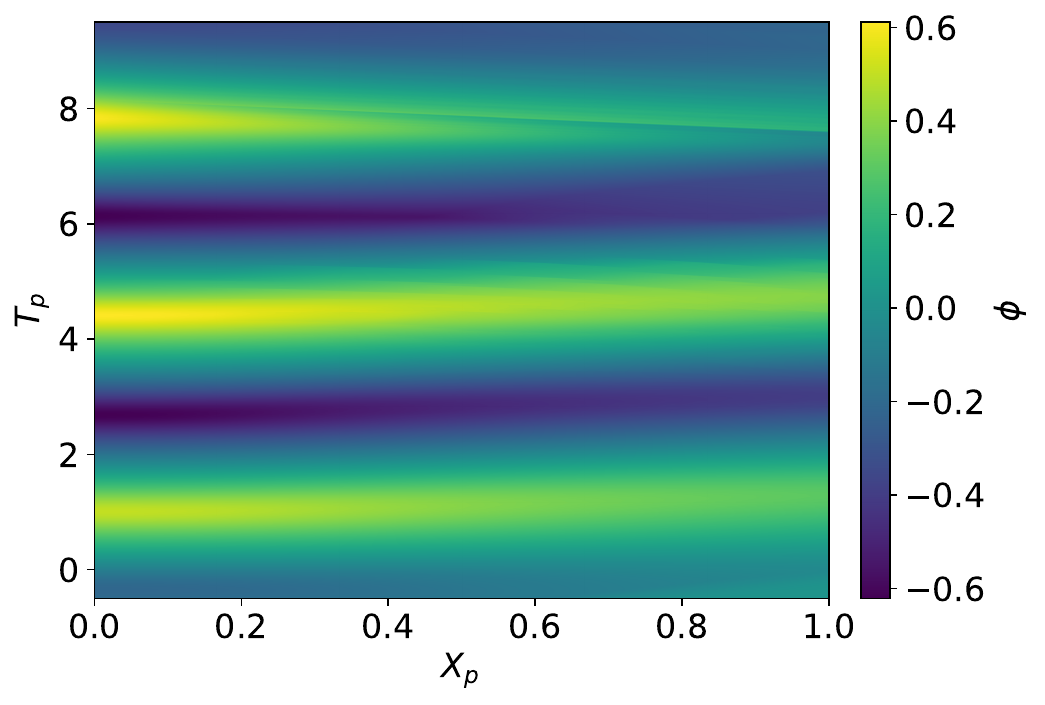}
  \caption{The scalar field is plotted as a function of slow time and
    the similarity coordinate in the $x$-direction $X_p$ as defined
    in~\eqref{eq:xi_x_prop}. These correspond to the highest
    subcritical evolutions of incoming data with the adjusted damping
    of Eq.~\eqref{eq:damp_hack} and on the left with
    HDWG-$\ln(\alpha)$ (configuration~\ref{id:Inc_g2oalpha_smoo_ln})
    and on the right with the DSS-compatible gauge source function
    (configuration \ref{id:Inc_g2oalpha_smoo_DSS}).}
  \label{fig:heat_maps}
 \end{figure*}
\begin{figure}[t]
  \includegraphics[width=\columnwidth]{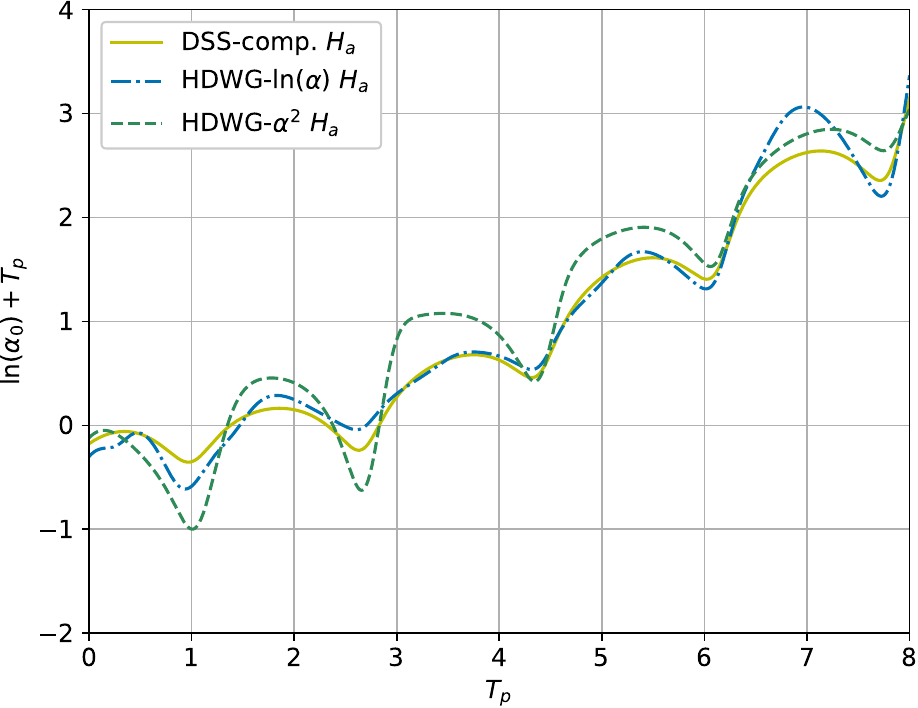}
  \includegraphics[width=\columnwidth]{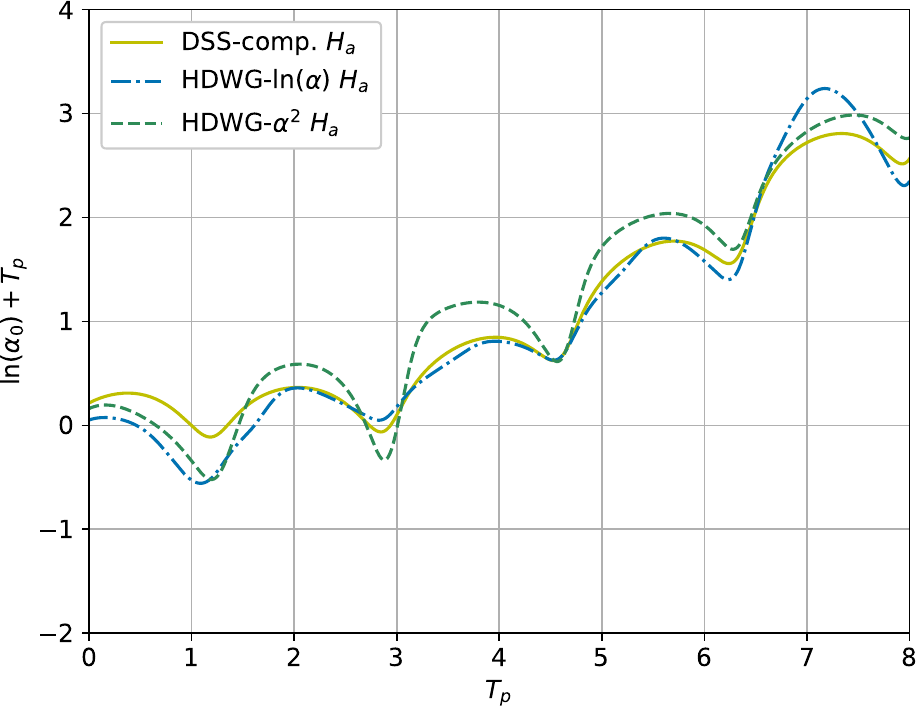}
  \caption{Left hand side of Eq.~\eqref{eq:ln_alpha_T} for the highest
    amplitude of bisections of incoming data on the top panel and
    moment of time symmetry data in the lower panel both with adjusted
    damping of Eq.~\eqref{eq:damp_hack} and both HDWG-variants~$H_a$
    (configurations~\ref{id:Inc_g2oalpha_smoo_ln}
    and~\ref{id:Inc_g2oalpha_smoo_alpha2}) and the
    DSS-compatible~$H_a$, Eq.~\eqref{eq:DSS_Ha} (configuration
    \ref{id:Inc_g2oalpha_smoo_DSS}). Despite the different data used,
    each curve seems identical in both panels, mostly determined by
    the choice of $H_a$.\label{fig:gsf_alpha_echoes_T}}
\end{figure}
\begin{figure}[t]
  \includegraphics[width=\columnwidth]{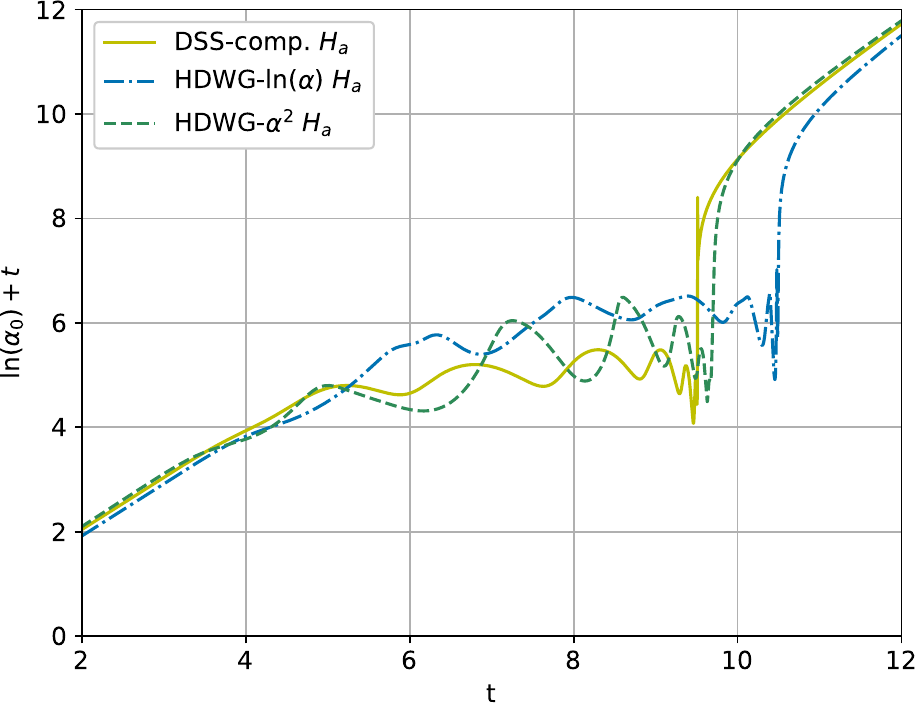}
  \includegraphics[width=\columnwidth]{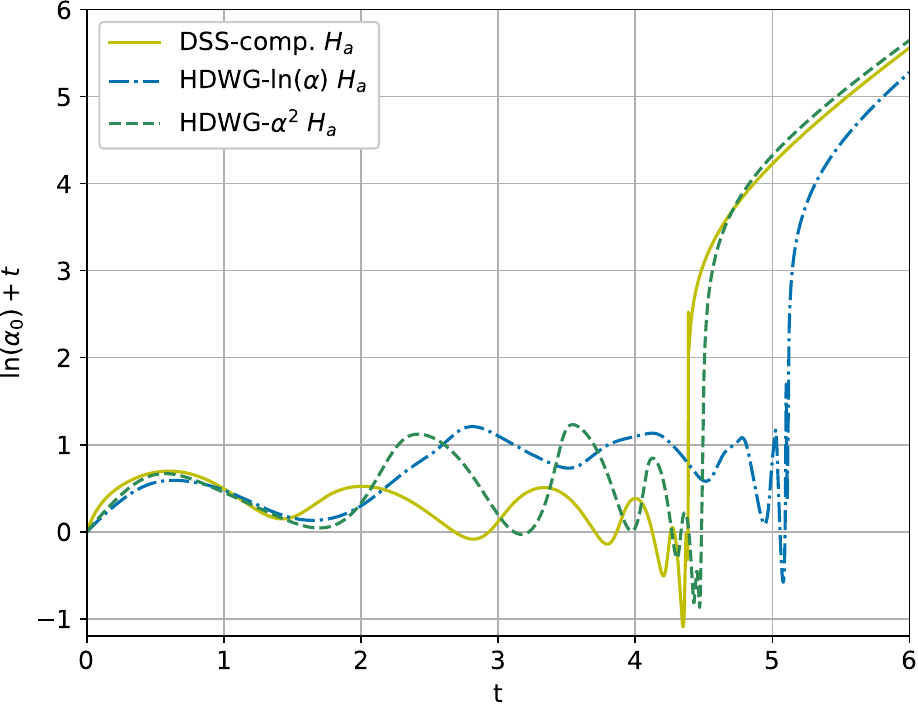}
  \caption{Left hand side of Eq.~\eqref{eq:ln_alpha_t} for the highest
    amplitude of bisections of incoming data on the top panel and
    moment of time symmetry data in the lower panel both with adjusted
    damping of Eq.~\eqref{eq:damp_hack} and both HDWG-variants~$H_a$
    (configurations~\ref{id:Inc_g2oalpha_smoo_ln}
    and~\ref{id:Inc_g2oalpha_smoo_alpha2}) and the
    DSS-compatible~$H_a$, Eq.~\eqref{eq:DSS_Ha} (configuration
    \ref{id:Inc_g2oalpha_smoo_DSS}). The top panel is zoomed out to
    show how the DSS phase stands out in the
    evolution.\label{fig:gsf_alpha_echoes_t} }
\end{figure}

As positive as it is to tackle the issues that were stopping our
critical collapse bisections, the question of how well any of our
gauges dynamically adapt to the approximate self-similarity of the
evolved spacetimes remains. We compare directly with the
HDWG-$\ln(\alpha)$ non-DSS-compatible gauge source function, because
we built the DSS-compatible suggestion so as to approximately agree
with it, as can be seen in Fig.~\ref{fig:slicing_Ha}, but also because
they provide very comparable results for critical phenomena, as
demonstrated in Fig.~\ref{fig:gsf_echoes}.

First, we examine the presence of DSS in the fields evolved. After the
evolution we transform the evolved coordinates into constructed
canonical DSS-adapted coordinates to verify the symmetry. These
coordinates are slow time computed from proper time~$T_p$ at the
center, as defined in Eq.~\eqref{eq:slow_time_prop}, and its spatial
counterpart
\begin{align}
  \label{eq:xi_x_prop}
  X_p^i = \frac{x_p^i}{|\tau_{*} - \tau|} \,,
\end{align}
where~$x_p^i$ are proper lengths. In Fig.~\ref{fig:heat_maps} we plot
the scalar field of a near-critical spacetime evolved with
HDWG-$\ln(\alpha)$ and the DSS-compatible gauge sources, with respect
to~$(T_p, X_p^i)$. This figure helps us also assess if any unwanted
gauge features actually disrupt the self-similarity phase of the
spacetime. This is luckily not the case for either gauge choice. It
seems that any undesirable gauge features reported in the previous
subsection~\ref{subsec:coord} are not sufficient to prevent reliable
bisections, and not enough to disrupt the spacetimes that survive. The
periodicity observed in these plots also confirms that we are close to
the threshold, the only case where the similarity would be exact.

We proceed to examine whether our evolved coordinates are adapted to
DSS. If our evolved time coordinate~$t$ was exactly~$T_p$, as defined
in Eq.~\eqref{eq:slow_time_prop}, then Eq.~\eqref{eq:DSS_alpha} would
hold, which can be rewritten in the same way as
Eq.~\eqref{eq:DSS_rescale} as
\begin{align}
  \label{eq:alpha_rescale}
  \alpha\left(T_p, X_p^i\right)
  &= e^{-T_p} \tilde{\alpha}\left(T_p, X_p^i\right) \, ,\\
  \label{eq:ln_alpha_T}
  \ln\left(\alpha\left(T_p, X_p^i\right)\right)
  &+ T_p
    = \ln\left(\tilde{\alpha}\left(T_p, X_p^i\right)\right) \, ,
\end{align}
where~$\tilde{\alpha}$ is~$\Delta$-periodic in~$T_p$ at fixed spatial
similarity coordinate. Fig.~\ref{fig:gsf_alpha_echoes_T} we plot the
left hand side of Eq.~\eqref{eq:ln_alpha_T} to check if indeed we find
a periodic function of constant average
as~$\ln\left(\tilde{\alpha}(T_p, X_p^i)\right)$ is. The periodicity is
certainly noticeable for all gauge source functions, but the
non-vanishing slope indicates that Eq.~\eqref{eq:ln_alpha_T} is not
satisfied for either gauge choice. This means that~$t \neq T_p$.

The DSS-adapted coordinates~$(T_p, X_p^i)$ are not unique. As it is
explained in Appendix~\ref{app:T}, any coordinates $(T,X^i)$ related
to~$(T_p, X_p^i)$ by
\begin{align}
	\label{eq:dss_coord_relation}
	T &= T_p + \tilde{f}^{T} \left(T_p, X_p^i\right) \, , \\
	X^i &=\tilde{f}^{i} \left(T_p, X_p^i\right)\, ,
\end{align}
where~$\tilde{f}^{a} \left(T_p, X_p^i\right)$ is a
function~$\Delta$-periodic in $T_p$, will be themselves DSS-adapted.
The existence of such a wide range of DSS-adapted coordinates could
make us hope that even if our coordinate $t$ does not exactly coincide
with~$T_p$, it could still coincide with $T$
in~\eqref{eq:dss_coord_relation} and hence be DSS-adapted. However, a
short calculation also provided in Appendix~\ref{app:T} shows that
if~$t$ was DSS-adapted it would satisfy the equation
\begin{align}
	\label{eq:ln_alpha_Tprop}
	\ln\left(\alpha\left(t, x^i\right)\right) + T_p = f \, ,
\end{align}
where~$f$ must be a function~$\Delta$-periodic in~$T_p$.
Fig.~\ref{fig:gsf_alpha_echoes_T} shows that~$\ln(\alpha) + T_p$ has
some oscillations with a characteristic frequency, but it is not
periodic in~$T_p$, which implies that~$t$ is not a DSS-adapted time
coordinate.
 
We can see more explicitly that~$t$ is not DSS-adapted at all. For our
coordinate~$t$ to be DSS-adapted we should have that
\begin{align}
  \label{eq:ln_alpha_t}
  \ln(\alpha(t, x^i)) + t = \ln\left(\tilde{\alpha}(t, x^i)\right) \, .
\end{align}
In Fig.~\ref{fig:gsf_alpha_echoes_t} we plot the left hand side of
Eq.~\eqref{eq:ln_alpha_t} to check again if we see a periodic function
of constant average.  We surprisingly observe opposite results to
those of Fig.~\ref{fig:gsf_alpha_echoes_T}: although the average
through a constant value during the self-similar phase is much more
noticeable than in Fig.~\ref{fig:gsf_alpha_echoes_T}, there is not
such a clear periodicty. We can now conclude that the dynamical
coordinates associated with neither gauge are adapted to
DSS. Following the calculation in Section~\ref{sec:dss} we already
knew that any non-DSS-compatible gauge choice could not yield
DSS-adapted coordinates, and that even a DSS-compatible one could not
guarantee them. The gauge choice Eq.~\eqref{eq:DSS_Ha} is
unfortunately not the needle in the haystack.

The choice we made for a DSS-compatible gauge source function is only
a first attempt. Many other gauge source functions could be
constructed that satisfy the DSS-compatibility condition. As we
already saw in Fig.~\ref{fig:slicing_Ha}, the DSS-compatible gauge
source function we suggested is qualitatively closest to
HDWG-$\ln(\alpha)$ gauge source function, which may suggest the reason
that this variant performs the best of all the non-DSS-compatible
choices.

\section{Summary and discussion}
\label{sec:summary}

In this paper, using the classic example of scalar field collapse as a
testbed~\cite{Cho93}, we have overcome two major obstacles that were
stopping our critical collapse simulations of Brill waves
in~\cite{SuaRenCor22}, namely constraint violations and undesirable
coordinate features.

Taking into account the collapse of the lapse in near-critical
spacetimes, we modified the reduction constraint damping parameter of
GHG, $\gamma_2$, such that it keeps damping constraint violations near
criticality. This can be thought of as changing the damping timescale
from proper to coordinate time, and has allowed us to improve our
critical parameter estimation and thus to clearly observe critical
phenomena with two different families of real massless scalar fields
in spherical symmetry minimally coupled to GR. Since pseudospectral
codes typically employ first order systems of equations, GHG in our
case, we suspected that the inefficient damping of violations of the
reduction constraint of GHG in spacetimes where the lapse collapses
was responsible for the poorer tuning provided by pseudospectral
searches of critical collapse in comparison to those performed with
finite difference codes and second order in space formulations of
GR. With the improved damping scheme we can now confidently tune to a
level competitive with bespoke spherical methods. We believe the
bottleneck in tuning now to be the effect of round-off error.

Aiming to exploit the structure of a DSS spacetime to avoid coordinate
singularities, we derived a necessary condition for a gauge choice to
be compatible with DSS. The condition is appropriate also for CSS
spacetimes. Using a specific gauge source function for GHG that
satisfies the DSS-compatibility condition, we have reproduced the
critical phenomena of a massless scalar field at the level of commonly
used gauge choices. The spatial profiles of the fields suggests that
this gauge source function respects DSS slightly better than the other
gauge sources, particularly vanilla HDWG, which was not designed with
critical collapse in mind (see~\cite{Cor23} for discussion). However,
at the level of tuning that can be achieved with double-precision
arithmetic, we do not see much improvement in the observation of
critical phenomena, constraint violation reduction nor in the
avoidance of coordinate singularities with the DSS-compatible choice
we made. This is more a practical than a principle issue. Every time
we extend from one scale echo to the next, a larger class of
coordinate singularities may occur if we do not satisfy the
compatibility condition. Further work is however needed to come up
with gauge source functions that give truly DSS-adapted coordinates.

The natural continuation of the present work is simply to apply our
improvements to the more interesting setup of axisymmetry, both in the
context of scalar fields and gravitational waves. We expect that an
improved apparent horizon finder will be needed to classify spacetimes
of the latter. By systematically treating each of the obstructions we
encounter, we expect to be able to further investigate non-spherical
spacetimes more accurately, with the ultimate aim of gaining a
comprehensive understanding of the threshold of black hole formation
in complete generality.

\acknowledgments

We would like to thank Florian Atteneder for all his help throughout
this project, and Isabel Suárez Fernández and Krinio Marouda for
helpful discussions and feedback. This work has been supported by the
Deutsche Forschungsgemeinschaft (DFG) under Grant No. 406116891 within
the Research Training Group RTG 2522/1. Hannes R. Rüter acknowledges
support from the Funda\c c\~ao para a Ci\^encia e Tecnologia (FCT)
within the projects UID/04564/2021, UIDB/04564/2020, UIDP/04564/2020
and EXPL/FIS-AST/0735/2021 and The H2020 ERC Advanced Grant “Black
holes: gravitational engines of discovery” grant agreement
no. Gravitas–101052587. This work was supported in part by FCT
(Portugal) Project No. UIDB/00099/2020.

\appendix
\section{Eigenvalue problem }
\label{app:eigen}

The left eigenvectors~$\mathbf{l}_{s}$ of~$\mathbf{M}$ in~\eqref{eq:M}
are
\begin{align}
  \mathbf{l}_{s_1}{}^{T} &=  \left( 
            \begin{matrix}
              \frac{-|\omega|(\alpha +\beta^{\hat{\omega}}\gamma_1)(|\omega| + i \gamma_2)}{\gamma_1\gamma_2} \\
              \frac{(|\omega|(\alpha +\beta^{\hat{\omega}}\gamma_1) + i\alpha \gamma_2)(|\omega| + i \gamma_2)}{\gamma_1\gamma_2{}^2} \\
              \frac{(|\omega|(\alpha +\beta^{\hat{\omega}}\gamma_1) - i\beta^{\hat{\omega}}\gamma_1 \gamma_2)(|\omega| + i \gamma_2)}{\gamma_1\gamma_2{}^2}\\
   			  \beta^{A} 
            \end{matrix}
  \right), \\
  \mathbf{l}_{s_2}{}^{T} &=  \left( 
            \begin{matrix}
              \frac{-|\omega|(\alpha -\beta^{\hat{\omega}}\gamma_1)(|\omega| - i \gamma_2)}{\gamma_1\gamma_2} \\
              \frac{(|\omega|(\alpha -\beta^{\hat{\omega}}\gamma_1) - i\alpha \gamma_2)(|\omega| - i \gamma_2)}{\gamma_1\gamma_2{}^2}\\
              \frac{(|\omega|(-\alpha +\beta^{\hat{\omega}}\gamma_1) +
              i\beta^{\hat{\omega}}\gamma_1\gamma_2)(|\omega| - i \gamma_2))}{\gamma_1\gamma_2{}^2}\\
              \beta^{A} 
            \end{matrix} 
  \right) \, ,\\
  \mathbf{l}_{s_3, s_4}{}^{T} &= \left( 
            \begin{matrix}
              0\\
              0\\
              0\\
              \delta^{B}{}_{A}
            \end{matrix} 
  \right) , \quad
  \mathbf{l}_{s_5}{}^{T} = \left( 
  \begin{matrix}
    -i|\omega|\beta^{\hat{\omega}}\\
    0\\
    \beta^{\hat{\omega}}\\
	\beta^{A}
  \end{matrix}
  \right)\, .
\end{align}
The characteristic-like variables are
\begin{align}
  v_{s_1} &=
            \frac{(|\omega| + i \gamma_2) (|\omega|(\alpha + \beta^{\hat{\omega}}\gamma_1) + i\alpha\gamma_2) \Pi}{\gamma_2} \nonumber\\*
          &\quad + \gamma_1\gamma_2\beta^{A}\Phi_{A}\nonumber\\*
          &\quad + \big(i |\omega| \alpha + \frac{|\omega|^2(\alpha + \beta^{\hat{\omega}}\gamma_1)}{\gamma_2} + \gamma_1\gamma_2\beta^{\hat{\omega}}\big)\Phi_{\hat{\omega}} \nonumber\\*
          &\quad -|\omega|(\alpha + \gamma_1\beta^{\hat{\omega}})(|\omega| - i \gamma_2) g \, , \\
  v_{s_2} &= 			
            \frac{(|\omega| - i \gamma_2) (|\omega|(\alpha - \beta^{\hat{\omega}}\gamma_1) - i\alpha\gamma_2) \Pi}{\gamma_2} \nonumber\\*
          &\quad + \gamma_1\gamma_2\beta^{A}\Phi_{A}\nonumber\\*
          &\quad + \big(i |\omega| \alpha + \frac{|\omega|^2(-\alpha + \beta^{\hat{\omega}}\gamma_1)}{\gamma_2} + \gamma_1\gamma_2\beta^{\hat{\omega}}\big)\Phi_{\hat{\omega}}\nonumber\\*
          &\quad +|\omega|(-\alpha + \gamma_1\beta^{\hat{\omega}})(|\omega| - i \gamma_2) g \, ,\\
  v_{s_4, s_3} &= 	\Phi_{A} \, ,\\
  v_{s_5} &= \beta^{A}\Phi_{A} +
            \beta^{\hat{\omega}}(\Phi_{\hat{\omega}} - i|\omega|g)\, .
\end{align}

\section{DSS-adapted time coordinates}
\label{app:T}

Consider a set of coordinates $x^{a}=\left( t, x^i \right)$ that are
not adapted to DSS. Also consider two different foliations
corresponding to two sets of DSS-adapted coordinates
$x^{\bar{a}}=\left(\bar{t}, x^{\bar{i}}\right)$ and
$x^{\hat{a}} =\left(\hat{t}, x^{\hat{i}}\right)$.

Let a spacetime enter the DSS phase and call echo $1$ the event where
the scalar field first reaches $0.6$ and echo $2$ the subsequent event
where the field reaches again $0.6$.  In general we have
$\left(t_2, x_2^i\right) = \left(t_1 + \delta t , x_1^i + \delta
  x^i\right)$ with arbitrary $\delta t$ and $\delta x^i$ whilst being
in DSS-adapted coordinates implies $\delta t = \Delta$ and
$\delta x^i = 0$.

Assuming that there is a functional relationship between the
coordinates such that~$x^{\bar{a}} = h^{\bar{a}} \left( x^{a}\right)$
and~$x^{\bar{a}} = f^{\bar{a}} \left( x^{\hat{a}}\right)$ then
\begin{align}
  \label{eq:app_no_DSS_coord}
  \bar{t}_2 &= h^{\bar{t}}\left(t_2, x_2^i\right) =h^{\bar{t}}\left(t_1 + \delta t, x_1^i\right) \nonumber\\
            &=\bar{t}_1 + \Delta =  h^{\bar{t}}\left(t_1, x_1^i\right) +\Delta \, , \\
  \label{eq:app_DSS_coord}
  \bar{t}_2 &= f^{\bar{t}}\left(\hat{t}_2, x_2^{\hat{i}}\right) =f^{\bar{t}}\left(\hat{t}_1 + \Delta, x_1^{\hat{i}}\right) \nonumber\\
            &=\bar{t}_1 + \Delta =  f^{\bar{t}}\left(\hat t_1, x_1^{\hat{i}}\right) + \Delta \, . 
\end{align}
From Eq.~\eqref{eq:app_DSS_coord} we can conclude
that~$f^{\bar{t}}\left(\hat{t}_1 + \Delta, x_1^{\hat{i}}\right) =
f^{\bar{t}}\left(\hat t_1, x_1^{\hat{i}}\right) + \Delta$ and hence
that the function
\begin{align}
  \label{eq:app_tbar_that}
  \tilde{f}^{\bar{t}} \left(\hat{t}, x^{\hat{i}}\right)
  := f^{\bar{t}}\left(\hat{t}, x^{\hat{i}}\right) - \hat{t} = \bar{t} - \hat{t}
\end{align}
must be~$\Delta$-periodic in~$\hat{t}$. Note that this is only the
case because the DSS period is the same for both DSS-adapted
coordinates, which is why it does not apply to the not-DSS-adapted
coordinates as Eq.~\eqref{eq:app_no_DSS_coord} shows. Similarly, we
have
that~$x^{\bar{i}} = \tilde{f}^{\bar{i}} \left(\hat{t},
  x^{\hat{i}}\right)$ is also $\Delta$-periodic in $\hat{t}$.  From
Eq.~\eqref{eq:app_tbar_that} we have that
\begin{align}
  \label{eq:app_covD_tbar}
  \nabla_{\bar{a}} \bar{t} = \frac{\partial x^{\hat{a}}}{\partial x^{\bar{a}}}
  \left[\nabla_{\hat{a}} \hat{t} +\nabla_{\hat{a}} \tilde{f}^{\bar{t}}
  \left(\hat{t}, x^{\hat{i}}\right)\right] =
  \frac{\partial x^{\hat{a}}}{\partial x^{\bar{a}}}\, \tilde{C}_{\hat{a}}\left(x^{\hat{b}}\right) \, , 
\end{align}
where everything inside the square brackets, now referred to
as~$\tilde{C}_{\hat{a}}\left(x^{\hat{b}}\right)$, is $\Delta$-periodic
in $\hat{t}$.

As in Eq.~\eqref{eq:alpha_rescale}, in these DSS-adapted coordinates
we have
\begin{align}
  \label{eq:app_alpha_rescaled_bar}
  \bar{\alpha} \left(\bar{t}, x^{\bar{i}}\right)
  &= e^{-\bar{t}}\tilde{\bar{\alpha}} \left(\bar{t}, x^{\bar{i}}\right) \, , \\
  \label{eq:app_alpha_rescaled_hat}
  \hat{\alpha} \left(\hat{t}, x^{\hat{i}}\right)
  &= e^{-\hat{t}}\tilde{\hat{\alpha}} \left(\hat{t}, x^{\hat{i}}\right) \, ,
\end{align}
where~$\tilde{\bar{\alpha}} \left(\bar{t}, x^{\bar{i}}\right)$
and~$\tilde{\hat{\alpha}}\left(\hat{t}, x^{\hat{i}}\right)$
are~$\Delta$-periodic in~$\bar{t}$ and~$\hat{t}$ respectively. In
terms of the conformal metrics~$\tilde{g}^{\bar{a}\bar{b}}$
and~$\tilde{g}^{\hat{a}\hat{b}}$, $\Delta$-periodic in~$\bar{t}$
and~$\hat{t}$ respectively, we have that Eq.~\eqref{eq:app_covD_tbar}
implies
\begin{align}
  \label{eq:app_periodic_alpha}
  \tilde{\bar{\alpha}} \left(\bar{t}, x^{\bar{i}}\right)
  &=\left(-\tilde{g}^{\bar{a}\bar{b}}\,
    \nabla_{\bar{a}}\bar{t}\,\nabla_{\bar{b}}\bar{t}\right)^{-1/2}\nonumber \\
  &=\left(-\tilde{g}^{\bar{a}\bar{b}}\,
    \left(\frac{\partial x^{\hat{a}}}{\partial x^{\bar{a}}}
    \tilde{C}_{\hat{a}}\left(x^{\hat{c}}\right)\right) \,
    \left(\frac{\partial x^{\hat{b}}}{\partial x^{\bar{b}}}
    \tilde{C}_{\hat{b}}\left(x^{\hat{d}}\right)\right)\right)^{-1/2} \nonumber \\
  &=\left(-\tilde{g}^{\hat{a}\hat{b}}\,
    \tilde{C}_{\hat{a}}\left(x^{\hat{c}}\right)\, \tilde{C}_{\hat{b}}\left(x^{\hat{d}}\right)\right)^{-1/2} \, ,
\end{align}
such that~$\tilde{\bar{\alpha}} \left(\bar{t}, x^{\bar{i}}\right)$ is
also~$\Delta$-periodic in~$\hat{t}$.

Lastly, taking the logarithm of Eq.~\eqref{eq:app_alpha_rescaled_bar}
we have that
\begin{align}
  \ln\left(\bar{\alpha} \left(\bar{t}, x^{\bar{i}}\right) \right) + \bar{t}
  &=\ln\left(\tilde{\bar{\alpha}} \left(\bar{t}, x^{\bar{i}}\right) \right) \, ,\\
  \label{eq:app_periodic_lnalpha}
  \ln\left(\bar{\alpha} \left(\bar{t}, x^{\bar{i}}\right) \right) + \hat{t}
  &=\ln\left(\tilde{\bar{\alpha}} \left(\bar{t}, x^{\bar{i}}\right) \right)
    - \tilde{f}^{\bar{t}} \left(\hat{t}, x^{\hat{i}}\right) \, ,
\end{align}
where the right hand side should be~$\Delta$-periodic in~$\hat{t}$ as
Eq.~\eqref{eq:app_periodic_alpha} shows.

For the coordinates in the main text, we computed slow time from
proper time~$T_p$ which corresponds to~$\hat{t}$ and we would like to
check whether our evolution coordinate $t$ corresponds to another
DSS-adapted coordinate here denoted as~$\bar{t}$.  The function~$f$ in
Eq.~\eqref{eq:ln_alpha_Tprop} corresponds to the right hand side of
Eq.~\eqref{eq:app_periodic_lnalpha}.

\bibliography{FICC.bbl}

\begin{thebibliography}{46}%
\makeatletter
\providecommand \@ifxundefined [1]{%
 \@ifx{#1\undefined}
}%
\providecommand \@ifnum [1]{%
 \ifnum #1\expandafter \@firstoftwo
 \else \expandafter \@secondoftwo
 \fi
}%
\providecommand \@ifx [1]{%
 \ifx #1\expandafter \@firstoftwo
 \else \expandafter \@secondoftwo
 \fi
}%
\providecommand \natexlab [1]{#1}%
\providecommand \enquote  [1]{``#1''}%
\providecommand \bibnamefont  [1]{#1}%
\providecommand \bibfnamefont [1]{#1}%
\providecommand \citenamefont [1]{#1}%
\providecommand \href@noop [0]{\@secondoftwo}%
\providecommand \href [0]{\begingroup \@sanitize@url \@href}%
\providecommand \@href[1]{\@@startlink{#1}\@@href}%
\providecommand \@@href[1]{\endgroup#1\@@endlink}%
\providecommand \@sanitize@url [0]{\catcode `\\12\catcode `\$12\catcode
  `\&12\catcode `\#12\catcode `\^12\catcode `\_12\catcode `\%12\relax}%
\providecommand \@@startlink[1]{}%
\providecommand \@@endlink[0]{}%
\providecommand \url  [0]{\begingroup\@sanitize@url \@url }%
\providecommand \@url [1]{\endgroup\@href {#1}{\urlprefix }}%
\providecommand \urlprefix  [0]{URL }%
\providecommand \Eprint [0]{\href }%
\providecommand \doibase [0]{https://doi.org/}%
\providecommand \selectlanguage [0]{\@gobble}%
\providecommand \bibinfo  [0]{\@secondoftwo}%
\providecommand \bibfield  [0]{\@secondoftwo}%
\providecommand \translation [1]{[#1]}%
\providecommand \BibitemOpen [0]{}%
\providecommand \bibitemStop [0]{}%
\providecommand \bibitemNoStop [0]{.\EOS\space}%
\providecommand \EOS [0]{\spacefactor3000\relax}%
\providecommand \BibitemShut  [1]{\csname bibitem#1\endcsname}%
\let\auto@bib@innerbib\@empty
\bibitem [{\citenamefont {Choptuik}(1993)}]{Cho93}%
  \BibitemOpen
  \bibfield  {author} {\bibinfo {author} {\bibfnamefont {M.~W.}\ \bibnamefont
  {Choptuik}},\ }\bibfield  {title} {\bibinfo {title} {Universality and scaling
  in gravitational collapse of a massless scalar field},\ }\href
  {https://doi.org/10.1103/PhysRevLett.70.9} {\bibfield  {journal} {\bibinfo
  {journal} {Phys. Rev. Lett.}\ }\textbf {\bibinfo {volume} {70}},\ \bibinfo
  {pages} {9} (\bibinfo {year} {1993})}\BibitemShut {NoStop}%
\bibitem [{\citenamefont {Gundlach}(1997)}]{Gun96a}%
  \BibitemOpen
  \bibfield  {author} {\bibinfo {author} {\bibfnamefont {C.}~\bibnamefont
  {Gundlach}},\ }\bibfield  {title} {\bibinfo {title} {Understanding critical
  collapse of a scalar field},\ }\href
  {https://doi.org/10.1103/PhysRevD.55.695} {\bibfield  {journal} {\bibinfo
  {journal} {Phys. Rev. D}\ }\textbf {\bibinfo {volume} {55}},\ \bibinfo
  {pages} {695} (\bibinfo {year} {1997})},\ \Eprint
  {https://arxiv.org/abs/gr-qc/9604019} {arXiv:gr-qc/9604019} \BibitemShut
  {NoStop}%
\bibitem [{\citenamefont {Hod}\ and\ \citenamefont {Piran}(1997)}]{HodPir97}%
  \BibitemOpen
  \bibfield  {author} {\bibinfo {author} {\bibfnamefont {S.}~\bibnamefont
  {Hod}}\ and\ \bibinfo {author} {\bibfnamefont {T.}~\bibnamefont {Piran}},\
  }\bibfield  {title} {\bibinfo {title} {{Fine structure of Choptuik's mass
  scaling relation}},\ }\href {https://doi.org/10.1103/PhysRevD.55.R440}
  {\bibfield  {journal} {\bibinfo  {journal} {Phys.\ Rev.\ D}\ }\textbf
  {\bibinfo {volume} {55}},\ \bibinfo {pages} {440} (\bibinfo {year} {1997})},\
  \Eprint {https://arxiv.org/abs/gr-qc/9606087} {arXiv:gr-qc/9606087}
  \BibitemShut {NoStop}%
\bibitem [{\citenamefont {Reiterer}\ and\ \citenamefont
  {Trubowitz}(2019)}]{ReiTru19}%
  \BibitemOpen
  \bibfield  {author} {\bibinfo {author} {\bibfnamefont {M.}~\bibnamefont
  {Reiterer}}\ and\ \bibinfo {author} {\bibfnamefont {E.}~\bibnamefont
  {Trubowitz}},\ }\bibfield  {title} {\bibinfo {title} {{Choptuik’s Critical
  Spacetime Exists}},\ }\href {https://doi.org/10.1007/s00220-019-03413-8}
  {\bibfield  {journal} {\bibinfo  {journal} {Commun. Math. Phys.}\ }\textbf
  {\bibinfo {volume} {368}},\ \bibinfo {pages} {143} (\bibinfo {year}
  {2019})},\ \Eprint {https://arxiv.org/abs/1203.3766} {arXiv:1203.3766
  [gr-qc]} \BibitemShut {NoStop}%
\bibitem [{\citenamefont {Gundlach}\ and\ \citenamefont
  {Mart{\'i}n-Garc{\'i}a}(2007)}]{GunGar07}%
  \BibitemOpen
  \bibfield  {author} {\bibinfo {author} {\bibfnamefont {C.}~\bibnamefont
  {Gundlach}}\ and\ \bibinfo {author} {\bibfnamefont {J.~M.}\ \bibnamefont
  {Mart{\'i}n-Garc{\'i}a}},\ }\bibfield  {title} {\bibinfo {title} {Critical
  phenomena in gravitational collapse},\ }\href
  {https://doi.org/10.12942/lrr-2007-5} {\bibfield  {journal} {\bibinfo
  {journal} {Living Reviews in Relativity}\ }\textbf {\bibinfo {volume} {10}},\
  \bibinfo {pages} {5} (\bibinfo {year} {2007})},\ \Eprint
  {https://arxiv.org/abs/0711.4620} {arXiv:0711.4620 [gr-qc]} \BibitemShut
  {NoStop}%
\bibitem [{\citenamefont {Garfinkle}\ and\ \citenamefont
  {Duncan}(1998)}]{GarDun98}%
  \BibitemOpen
  \bibfield  {author} {\bibinfo {author} {\bibfnamefont {D.}~\bibnamefont
  {Garfinkle}}\ and\ \bibinfo {author} {\bibfnamefont {G.~C.}\ \bibnamefont
  {Duncan}},\ }\bibfield  {title} {\bibinfo {title} {{Scaling of curvature in
  subcritical gravitational collapse}},\ }\href
  {https://doi.org/10.1103/PhysRevD.58.064024} {\bibfield  {journal} {\bibinfo
  {journal} {Phys.Rev.}\ }\textbf {\bibinfo {volume} {D58}},\ \bibinfo {pages}
  {064024} (\bibinfo {year} {1998})},\ \Eprint
  {https://arxiv.org/abs/gr-qc/9802061} {arXiv:gr-qc/9802061 [gr-qc]}
  \BibitemShut {NoStop}%
\bibitem [{\citenamefont {Martín-García}\ and\ \citenamefont
  {Gundlach}(2003)}]{GarGun03}%
  \BibitemOpen
  \bibfield  {author} {\bibinfo {author} {\bibfnamefont {J.~M.}\ \bibnamefont
  {Martín-García}}\ and\ \bibinfo {author} {\bibfnamefont {C.}~\bibnamefont
  {Gundlach}},\ }\bibfield  {title} {\bibinfo {title} {Global structure of
  choptuik's critical solution in scalar field collapse},\ }\href
  {https://doi.org/10.1103/PhysRevD.68.024011} {\bibfield  {journal} {\bibinfo
  {journal} {Phys. Rev. D}\ }\textbf {\bibinfo {volume} {68}},\ \bibinfo
  {pages} {024011} (\bibinfo {year} {2003})},\ \Eprint
  {https://arxiv.org/abs/gr-qc/0304070} {arXiv:gr-qc/0304070} \BibitemShut
  {NoStop}%
\bibitem [{\citenamefont {Martin-Garcia}\ and\ \citenamefont
  {Gundlach}(1999)}]{GarGun98}%
  \BibitemOpen
  \bibfield  {author} {\bibinfo {author} {\bibfnamefont {J.~M.}\ \bibnamefont
  {Martin-Garcia}}\ and\ \bibinfo {author} {\bibfnamefont {C.}~\bibnamefont
  {Gundlach}},\ }\bibfield  {title} {\bibinfo {title} {All nonspherical
  perturbations of the choptuik spacetime decay},\ }\href
  {https://doi.org/10.1103/PhysRevD.59.064031} {\bibfield  {journal} {\bibinfo
  {journal} {Phys. Rev. D}\ }\textbf {\bibinfo {volume} {59}},\ \bibinfo
  {pages} {064031} (\bibinfo {year} {1999})},\ \Eprint
  {https://arxiv.org/abs/gr-qc/9809059} {arXiv:gr-qc/9809059} \BibitemShut
  {NoStop}%
\bibitem [{\citenamefont {Choptuik}\ \emph {et~al.}(2003)\citenamefont
  {Choptuik}, \citenamefont {Hirschmann}, \citenamefont {Liebling},\ and\
  \citenamefont {Pretorius}}]{ChoHirLie03}%
  \BibitemOpen
  \bibfield  {author} {\bibinfo {author} {\bibfnamefont {M.~W.}\ \bibnamefont
  {Choptuik}}, \bibinfo {author} {\bibfnamefont {E.~W.}\ \bibnamefont
  {Hirschmann}}, \bibinfo {author} {\bibfnamefont {S.~L.}\ \bibnamefont
  {Liebling}},\ and\ \bibinfo {author} {\bibfnamefont {F.}~\bibnamefont
  {Pretorius}},\ }\bibfield  {title} {\bibinfo {title} {Critical collapse of
  the massless scalar field in axisymmetry},\ }\href
  {https://doi.org/10.1103/PhysRevD.68.044007} {\bibfield  {journal} {\bibinfo
  {journal} {Phys. Rev. D}\ }\textbf {\bibinfo {volume} {68}},\ \bibinfo
  {pages} {044007} (\bibinfo {year} {2003})},\ \Eprint
  {https://arxiv.org/abs/gr-qc/0305003} {arXiv:gr-qc/0305003} \BibitemShut
  {NoStop}%
\bibitem [{\citenamefont {Baumgarte}(2018)}]{Bau18}%
  \BibitemOpen
  \bibfield  {author} {\bibinfo {author} {\bibfnamefont {T.~W.}\ \bibnamefont
  {Baumgarte}},\ }\bibfield  {title} {\bibinfo {title} {Aspherical deformations
  of the choptuik spacetime},\ }\href
  {https://doi.org/10.1103/physrevd.98.084012} {\bibfield  {journal} {\bibinfo
  {journal} {Phys. Rev. D}\ }\textbf {\bibinfo {volume} {98}},\ \bibinfo
  {pages} {084012} (\bibinfo {year} {2018})},\ \Eprint
  {https://arxiv.org/abs/1807.10342} {arXiv:1807.10342 [gr-qc]} \BibitemShut
  {NoStop}%
\bibitem [{\citenamefont {Healy}\ and\ \citenamefont
  {Laguna}(2014)}]{HeaLag13}%
  \BibitemOpen
  \bibfield  {author} {\bibinfo {author} {\bibfnamefont {J.}~\bibnamefont
  {Healy}}\ and\ \bibinfo {author} {\bibfnamefont {P.}~\bibnamefont {Laguna}},\
  }\bibfield  {title} {\bibinfo {title} {{Critical Collapse of Scalar Fields
  Beyond Axisymmetry}},\ }\href {https://doi.org/10.1007/s10714-014-1722-2}
  {\bibfield  {journal} {\bibinfo  {journal} {Gen. Rel. Grav.}\ }\textbf
  {\bibinfo {volume} {46}},\ \bibinfo {pages} {1722} (\bibinfo {year}
  {2014})},\ \Eprint {https://arxiv.org/abs/1310.1955} {arXiv:1310.1955
  [gr-qc]} \BibitemShut {NoStop}%
\bibitem [{\citenamefont {Deppe}\ \emph {et~al.}(2019)\citenamefont {Deppe},
  \citenamefont {Kidder}, \citenamefont {Scheel},\ and\ \citenamefont
  {Teukolsky}}]{DepKidSch18}%
  \BibitemOpen
  \bibfield  {author} {\bibinfo {author} {\bibfnamefont {N.}~\bibnamefont
  {Deppe}}, \bibinfo {author} {\bibfnamefont {L.~E.}\ \bibnamefont {Kidder}},
  \bibinfo {author} {\bibfnamefont {M.~A.}\ \bibnamefont {Scheel}},\ and\
  \bibinfo {author} {\bibfnamefont {S.~A.}\ \bibnamefont {Teukolsky}},\
  }\bibfield  {title} {\bibinfo {title} {Critical behavior in 3d gravitational
  collapse of massless scalar fields},\ }\href
  {https://doi.org/10.1103/PhysRevD.99.024018} {\bibfield  {journal} {\bibinfo
  {journal} {Phys. Rev. D}\ }\textbf {\bibinfo {volume} {99}},\ \bibinfo
  {pages} {024018} (\bibinfo {year} {2019})},\ \Eprint
  {https://arxiv.org/abs/1802.08682} {arXiv:1802.08682 [gr-qc]} \BibitemShut
  {NoStop}%
\bibitem [{\citenamefont {Hilditch}\ \emph {et~al.}(2017)\citenamefont
  {Hilditch}, \citenamefont {Weyhausen},\ and\ \citenamefont
  {Br{\"u}gmann}}]{HilWeyBru17}%
  \BibitemOpen
  \bibfield  {author} {\bibinfo {author} {\bibfnamefont {D.}~\bibnamefont
  {Hilditch}}, \bibinfo {author} {\bibfnamefont {A.}~\bibnamefont
  {Weyhausen}},\ and\ \bibinfo {author} {\bibfnamefont {B.}~\bibnamefont
  {Br{\"u}gmann}},\ }\bibfield  {title} {\bibinfo {title} {{Evolutions of
  centered Brill waves with a pseudospectral method}},\ }\href
  {https://doi.org/10.1103/PhysRevD.96.104051} {\bibfield  {journal} {\bibinfo
  {journal} {Phys. Rev.}\ }\textbf {\bibinfo {volume} {D96}},\ \bibinfo {pages}
  {104051} (\bibinfo {year} {2017})},\ \Eprint
  {https://arxiv.org/abs/1706.01829} {arXiv:1706.01829 [gr-qc]} \BibitemShut
  {NoStop}%
\bibitem [{\citenamefont {Khirnov}\ and\ \citenamefont
  {Ledvinka}(2018)}]{KhiLed18}%
  \BibitemOpen
  \bibfield  {author} {\bibinfo {author} {\bibfnamefont {A.}~\bibnamefont
  {Khirnov}}\ and\ \bibinfo {author} {\bibfnamefont {T.}~\bibnamefont
  {Ledvinka}},\ }\bibfield  {title} {\bibinfo {title} {{Slicing conditions for
  axisymmetric gravitational collapse of Brill waves}},\ }\href
  {https://doi.org/10.1088/1361-6382/aae1bc} {\bibfield  {journal} {\bibinfo
  {journal} {Class. Quantum Grav.}\ }\textbf {\bibinfo {volume} {35}},\
  \bibinfo {pages} {215003} (\bibinfo {year} {2018})},\ \Eprint
  {https://arxiv.org/abs/1908.06034} {arXiv:1908.06034 [gr-qc]} \BibitemShut
  {NoStop}%
\bibitem [{\citenamefont {Ledvinka}\ and\ \citenamefont
  {Khirnov}(2021)}]{LedKhi21}%
  \BibitemOpen
  \bibfield  {author} {\bibinfo {author} {\bibfnamefont {T.}~\bibnamefont
  {Ledvinka}}\ and\ \bibinfo {author} {\bibfnamefont {A.}~\bibnamefont
  {Khirnov}},\ }\bibfield  {title} {\bibinfo {title} {Universality of curvature
  invariants in critical vacuum gravitational collapse},\ }\href
  {https://doi.org/10.1103/PhysRevLett.127.011104} {\bibfield  {journal}
  {\bibinfo  {journal} {Phys. Rev. Lett.}\ }\textbf {\bibinfo {volume} {127}},\
  \bibinfo {pages} {011104} (\bibinfo {year} {2021})},\ \Eprint
  {https://arxiv.org/abs/2102.09579} {arXiv:2102.09579 [gr-qc]} \BibitemShut
  {NoStop}%
\bibitem [{\citenamefont {Su\'arez~Fern\'andez}\ \emph
  {et~al.}(2022)\citenamefont {Su\'arez~Fern\'andez}, \citenamefont {Renkhoff},
  \citenamefont {Cors}, \citenamefont {Brügmann},\ and\ \citenamefont
  {Hilditch}}]{SuaRenCor22}%
  \BibitemOpen
  \bibfield  {author} {\bibinfo {author} {\bibfnamefont {I.}~\bibnamefont
  {Su\'arez~Fern\'andez}}, \bibinfo {author} {\bibfnamefont {S.}~\bibnamefont
  {Renkhoff}}, \bibinfo {author} {\bibfnamefont {D.}~\bibnamefont {Cors}},
  \bibinfo {author} {\bibfnamefont {B.}~\bibnamefont {Brügmann}},\ and\
  \bibinfo {author} {\bibfnamefont {D.}~\bibnamefont {Hilditch}},\ }\bibfield
  {title} {\bibinfo {title} {Evolution of brill waves with an adaptive
  pseudospectral method},\ }\href {https://doi.org/10.1103/PhysRevD.106.024036}
  {\bibfield  {journal} {\bibinfo  {journal} {Phys. Rev. D}\ }\textbf {\bibinfo
  {volume} {106}},\ \bibinfo {pages} {024036} (\bibinfo {year} {2022})},\
  \Eprint {https://arxiv.org/abs/2205.04379} {arXiv:2205.04379 [gr-qc]}
  \BibitemShut {NoStop}%
\bibitem [{\citenamefont {Baumgarte}\ \emph
  {et~al.}(2023{\natexlab{a}})\citenamefont {Baumgarte}, \citenamefont
  {Gundlach},\ and\ \citenamefont {Hilditch}}]{BauGunHil23}%
  \BibitemOpen
  \bibfield  {author} {\bibinfo {author} {\bibfnamefont {T.~W.}\ \bibnamefont
  {Baumgarte}}, \bibinfo {author} {\bibfnamefont {C.}~\bibnamefont
  {Gundlach}},\ and\ \bibinfo {author} {\bibfnamefont {D.}~\bibnamefont
  {Hilditch}},\ }\bibfield  {title} {\bibinfo {title} {{Critical phenomena in
  the collapse of quadrupolar and hexadecapolar gravitational waves}},\ }\href
  {https://doi.org/10.1103/PhysRevD.107.084012} {\bibfield  {journal} {\bibinfo
   {journal} {Phys. Rev. D}\ }\textbf {\bibinfo {volume} {107}},\ \bibinfo
  {pages} {084012} (\bibinfo {year} {2023}{\natexlab{a}})},\ \Eprint
  {https://arxiv.org/abs/2303.05530} {arXiv:2303.05530 [gr-qc]} \BibitemShut
  {NoStop}%
\bibitem [{\citenamefont {Baumgarte}\ \emph
  {et~al.}(2023{\natexlab{b}})\citenamefont {Baumgarte}, \citenamefont
  {Br\"ugmann}, \citenamefont {Cors}, \citenamefont {Gundlach}, \citenamefont
  {Hilditch}, \citenamefont {Khirnov}, \citenamefont {Ledvinka}, \citenamefont
  {Renkhoff},\ and\ \citenamefont {Fern\'andez}}]{BauBruCor23}%
  \BibitemOpen
  \bibfield  {author} {\bibinfo {author} {\bibfnamefont {T.~W.}\ \bibnamefont
  {Baumgarte}}, \bibinfo {author} {\bibfnamefont {B.}~\bibnamefont
  {Br\"ugmann}}, \bibinfo {author} {\bibfnamefont {D.}~\bibnamefont {Cors}},
  \bibinfo {author} {\bibfnamefont {C.}~\bibnamefont {Gundlach}}, \bibinfo
  {author} {\bibfnamefont {D.}~\bibnamefont {Hilditch}}, \bibinfo {author}
  {\bibfnamefont {A.}~\bibnamefont {Khirnov}}, \bibinfo {author} {\bibfnamefont
  {T.}~\bibnamefont {Ledvinka}}, \bibinfo {author} {\bibfnamefont
  {S.}~\bibnamefont {Renkhoff}},\ and\ \bibinfo {author} {\bibfnamefont
  {I.~S.}\ \bibnamefont {Fern\'andez}},\ }\href@noop {} {\bibinfo {title}
  {{Critical phenomena in the collapse of gravitational waves}}} (\bibinfo
  {year} {2023}{\natexlab{b}}),\ \Eprint {https://arxiv.org/abs/2305.17171}
  {arXiv:2305.17171 [gr-qc]} \BibitemShut {NoStop}%
\bibitem [{\citenamefont {Baumgarte}\ \emph {et~al.}(2019)\citenamefont
  {Baumgarte}, \citenamefont {Gundlach},\ and\ \citenamefont
  {Hilditch}}]{BauGunHil19}%
  \BibitemOpen
  \bibfield  {author} {\bibinfo {author} {\bibfnamefont {T.~W.}\ \bibnamefont
  {Baumgarte}}, \bibinfo {author} {\bibfnamefont {C.}~\bibnamefont
  {Gundlach}},\ and\ \bibinfo {author} {\bibfnamefont {D.}~\bibnamefont
  {Hilditch}},\ }\bibfield  {title} {\bibinfo {title} {{Critical phenomena in
  the gravitational collapse of electromagnetic waves}},\ }\href
  {https://doi.org/10.1103/PhysRevLett.123.171103} {\bibfield  {journal}
  {\bibinfo  {journal} {Phys. Rev. Lett.}\ }\textbf {\bibinfo {volume} {123}},\
  \bibinfo {pages} {171103} (\bibinfo {year} {2019})},\ \Eprint
  {https://arxiv.org/abs/1909.00850} {arXiv:1909.00850 [gr-qc]} \BibitemShut
  {NoStop}%
\bibitem [{\citenamefont {Perez~Mendoza}\ and\ \citenamefont
  {Baumgarte}(2021)}]{MenBau21}%
  \BibitemOpen
  \bibfield  {author} {\bibinfo {author} {\bibfnamefont {M.~F.}\ \bibnamefont
  {Perez~Mendoza}}\ and\ \bibinfo {author} {\bibfnamefont {T.~W.}\ \bibnamefont
  {Baumgarte}},\ }\bibfield  {title} {\bibinfo {title} {Critical phenomena in
  the gravitational collapse of electromagnetic dipole and quadrupole waves},\
  }\href {https://doi.org/10.1103/PhysRevD.103.124048} {\bibfield  {journal}
  {\bibinfo  {journal} {Phys. Rev. D}\ }\textbf {\bibinfo {volume} {103}},\
  \bibinfo {pages} {124048} (\bibinfo {year} {2021})},\ \Eprint
  {https://arxiv.org/abs/2104.03980} {arXiv:2104.03980 [gr-qc]} \BibitemShut
  {NoStop}%
\bibitem [{\citenamefont {Renkhoff}()}]{ahloc3d}%
  \BibitemOpen
  \bibfield  {author} {\bibinfo {author} {\bibfnamefont {S.}~\bibnamefont
  {Renkhoff}},\ }\href@noop {} {}\bibinfo {note}
  {\url{https://git.tpi.uni-jena.de/srenkhoff/ahloc3d}}\BibitemShut {NoStop}%
\bibitem [{\citenamefont {Alcubierre}(1997)}]{Alc96}%
  \BibitemOpen
  \bibfield  {author} {\bibinfo {author} {\bibfnamefont {M.}~\bibnamefont
  {Alcubierre}},\ }\bibfield  {title} {\bibinfo {title} {Appearance of
  coordinate shocks in hyperbolic formalisms of general relativity},\ }\href
  {https://doi.org/10.1103/PhysRevD.55.5981} {\bibfield  {journal} {\bibinfo
  {journal} {Phys. Rev. D}\ }\textbf {\bibinfo {volume} {55}},\ \bibinfo
  {pages} {5981} (\bibinfo {year} {1997})},\ \Eprint
  {https://arxiv.org/abs/gr-qc/9609015} {arXiv:gr-qc/9609015} \BibitemShut
  {NoStop}%
\bibitem [{\citenamefont {Alcubierre}(2003)}]{Alc02}%
  \BibitemOpen
  \bibfield  {author} {\bibinfo {author} {\bibfnamefont {M.}~\bibnamefont
  {Alcubierre}},\ }\bibfield  {title} {\bibinfo {title} {Hyperbolic slicings of
  spacetime: singularity avoidance and gauge shocks},\ }\href
  {https://doi.org/10.1088/0264-9381/20/4/304} {\bibfield  {journal} {\bibinfo
  {journal} {Class. Quantum Grav.}\ }\textbf {\bibinfo {volume} {20}},\
  \bibinfo {pages} {607} (\bibinfo {year} {2003})},\ \Eprint
  {https://arxiv.org/abs/gr-qc/0210050} {arXiv:gr-qc/0210050} \BibitemShut
  {NoStop}%
\bibitem [{\citenamefont {Baumgarte}\ and\ \citenamefont
  {Hilditch}(2022)}]{BauHil22}%
  \BibitemOpen
  \bibfield  {author} {\bibinfo {author} {\bibfnamefont {T.~W.}\ \bibnamefont
  {Baumgarte}}\ and\ \bibinfo {author} {\bibfnamefont {D.}~\bibnamefont
  {Hilditch}},\ }\bibfield  {title} {\bibinfo {title} {{Shock-avoiding slicing
  conditions: Tests and calibrations}},\ }\href
  {https://doi.org/10.1103/PhysRevD.106.044014} {\bibfield  {journal} {\bibinfo
   {journal} {Phys. Rev. D}\ }\textbf {\bibinfo {volume} {106}},\ \bibinfo
  {pages} {044014} (\bibinfo {year} {2022})},\ \Eprint
  {https://arxiv.org/abs/2207.06376} {arXiv:2207.06376 [gr-qc]} \BibitemShut
  {NoStop}%
\bibitem [{\citenamefont {Lindblom}\ and\ \citenamefont
  {Szil{\'a}gyi}(2009)}]{LinSzi09}%
  \BibitemOpen
  \bibfield  {author} {\bibinfo {author} {\bibfnamefont {L.}~\bibnamefont
  {Lindblom}}\ and\ \bibinfo {author} {\bibfnamefont {B.}~\bibnamefont
  {Szil{\'a}gyi}},\ }\bibfield  {title} {\bibinfo {title} {{An Improved Gauge
  Driver for the GH Einstein System}},\ }\href
  {https://doi.org/10.1103/PhysRevD.80.084019} {\bibfield  {journal} {\bibinfo
  {journal} {Phys. Rev.}\ }\textbf {\bibinfo {volume} {D80}},\ \bibinfo {pages}
  {084019} (\bibinfo {year} {2009})},\ \Eprint
  {https://arxiv.org/abs/0904.4873} {arXiv:0904.4873 [gr-qc]} \BibitemShut
  {NoStop}%
\bibitem [{\citenamefont {Garfinkle}\ and\ \citenamefont
  {Gundlach}(1999)}]{GarGun99}%
  \BibitemOpen
  \bibfield  {author} {\bibinfo {author} {\bibfnamefont {D.}~\bibnamefont
  {Garfinkle}}\ and\ \bibinfo {author} {\bibfnamefont {C.}~\bibnamefont
  {Gundlach}},\ }\bibfield  {title} {\bibinfo {title} {Symmetry-seeking
  spacetime coordinates},\ }\href {https://doi.org/10.1088/0264-9381/16/12/325}
  {\bibfield  {journal} {\bibinfo  {journal} {Classical and Quantum Gravity}\
  }\textbf {\bibinfo {volume} {16}},\ \bibinfo {pages} {4111} (\bibinfo {year}
  {1999})},\ \Eprint {https://arxiv.org/abs/gr-qc/9908016}
  {arXiv:gr-qc/9908016} \BibitemShut {NoStop}%
\bibitem [{\citenamefont {Hilditch}\ \emph {et~al.}(2016)\citenamefont
  {Hilditch}, \citenamefont {Weyhausen},\ and\ \citenamefont
  {Br{\"u}gmann}}]{HilWeyBru15}%
  \BibitemOpen
  \bibfield  {author} {\bibinfo {author} {\bibfnamefont {D.}~\bibnamefont
  {Hilditch}}, \bibinfo {author} {\bibfnamefont {A.}~\bibnamefont
  {Weyhausen}},\ and\ \bibinfo {author} {\bibfnamefont {B.}~\bibnamefont
  {Br{\"u}gmann}},\ }\bibfield  {title} {\bibinfo {title} {{Pseudospectral
  method for gravitational wave collapse}},\ }\href
  {https://doi.org/10.1103/PhysRevD.93.063006} {\bibfield  {journal} {\bibinfo
  {journal} {Phys. Rev.}\ }\textbf {\bibinfo {volume} {D93}},\ \bibinfo {pages}
  {063006} (\bibinfo {year} {2016})},\ \Eprint
  {https://arxiv.org/abs/1504.04732} {arXiv:1504.04732 [gr-qc]} \BibitemShut
  {NoStop}%
\bibitem [{\citenamefont {Renkhoff}\ \emph {et~al.}(2023)\citenamefont
  {Renkhoff}, \citenamefont {Cors}, \citenamefont {Hilditch},\ and\
  \citenamefont {Br\"ugmann}}]{RenCorHil23}%
  \BibitemOpen
  \bibfield  {author} {\bibinfo {author} {\bibfnamefont {S.}~\bibnamefont
  {Renkhoff}}, \bibinfo {author} {\bibfnamefont {D.}~\bibnamefont {Cors}},
  \bibinfo {author} {\bibfnamefont {D.}~\bibnamefont {Hilditch}},\ and\
  \bibinfo {author} {\bibfnamefont {B.}~\bibnamefont {Br\"ugmann}},\ }\bibfield
   {title} {\bibinfo {title} {{Adaptive hp refinement for spectral elements in
  numerical relativity}},\ }\href {https://doi.org/10.1103/PhysRevD.107.104043}
  {\bibfield  {journal} {\bibinfo  {journal} {Phys. Rev. D}\ }\textbf {\bibinfo
  {volume} {107}},\ \bibinfo {pages} {104043} (\bibinfo {year} {2023})},\
  \Eprint {https://arxiv.org/abs/2302.00575} {arXiv:2302.00575 [gr-qc]}
  \BibitemShut {NoStop}%
\bibitem [{\citenamefont {Rinne}(2006)}]{Rin06a}%
  \BibitemOpen
  \bibfield  {author} {\bibinfo {author} {\bibfnamefont {O.}~\bibnamefont
  {Rinne}},\ }\bibfield  {title} {\bibinfo {title} {{Stable
  radiation-controlling boundary conditions for the generalized harmonic
  Einstein equations}},\ }\href {https://doi.org/10.1088/0264-9381/23/22/013}
  {\bibfield  {journal} {\bibinfo  {journal} {Class. Quantum Grav.}\ }\textbf
  {\bibinfo {volume} {23}},\ \bibinfo {pages} {6275} (\bibinfo {year}
  {2006})},\ \Eprint {https://arxiv.org/abs/gr-qc/0606053}
  {arXiv:gr-qc/0606053} \BibitemShut {NoStop}%
\bibitem [{\citenamefont {R{\"u}ter}\ \emph {et~al.}(2018)\citenamefont
  {R{\"u}ter}, \citenamefont {Hilditch}, \citenamefont {Bugner},\ and\
  \citenamefont {Br{\"u}gmann}}]{RueHilBug17}%
  \BibitemOpen
  \bibfield  {author} {\bibinfo {author} {\bibfnamefont {H.~R.}\ \bibnamefont
  {R{\"u}ter}}, \bibinfo {author} {\bibfnamefont {D.}~\bibnamefont {Hilditch}},
  \bibinfo {author} {\bibfnamefont {M.}~\bibnamefont {Bugner}},\ and\ \bibinfo
  {author} {\bibfnamefont {B.}~\bibnamefont {Br{\"u}gmann}},\ }\bibfield
  {title} {\bibinfo {title} {{Hyperbolic Relaxation Method for Elliptic
  Equations}},\ }\href {https://doi.org/10.1103/PhysRevD.98.084044} {\bibfield
  {journal} {\bibinfo  {journal} {Phys. Rev.}\ }\textbf {\bibinfo {volume}
  {D98}},\ \bibinfo {pages} {084044} (\bibinfo {year} {2018})},\ \Eprint
  {https://arxiv.org/abs/1708.07358} {arXiv:1708.07358 [gr-qc]} \BibitemShut
  {NoStop}%
\bibitem [{\citenamefont {Baumgarte}\ and\ \citenamefont
  {Shapiro}(2010)}]{BauSha10}%
  \BibitemOpen
  \bibfield  {author} {\bibinfo {author} {\bibfnamefont {T.~W.}\ \bibnamefont
  {Baumgarte}}\ and\ \bibinfo {author} {\bibfnamefont {S.~L.}\ \bibnamefont
  {Shapiro}},\ }\href@noop {} {\emph {\bibinfo {title} {Numerical Relativity:
  Solving {E}instein's Equations on the Computer}}}\ (\bibinfo  {publisher}
  {Cambridge University Press},\ \bibinfo {address} {Cambridge},\ \bibinfo
  {year} {2010})\BibitemShut {NoStop}%
\bibitem [{\citenamefont {Tichy}(2017)}]{Tic17}%
  \BibitemOpen
  \bibfield  {author} {\bibinfo {author} {\bibfnamefont {W.}~\bibnamefont
  {Tichy}},\ }\bibfield  {title} {\bibinfo {title} {The initial value problem
  as it relates to numerical relativity},\ }\href
  {http://stacks.iop.org/0034-4885/80/i=2/a=026901} {\bibfield  {journal}
  {\bibinfo  {journal} {Reports on Progress in Physics}\ }\textbf {\bibinfo
  {volume} {80}},\ \bibinfo {pages} {026901} (\bibinfo {year} {2017})},\
  \Eprint {https://arxiv.org/abs/1610.03805} {arXiv:1610.03805 [gr-qc]}
  \BibitemShut {NoStop}%
\bibitem [{\citenamefont {Lindblom}\ \emph {et~al.}(2006)\citenamefont
  {Lindblom}, \citenamefont {Scheel}, \citenamefont {Kidder}, \citenamefont
  {Owen},\ and\ \citenamefont {Rinne}}]{LinSchKid05}%
  \BibitemOpen
  \bibfield  {author} {\bibinfo {author} {\bibfnamefont {L.}~\bibnamefont
  {Lindblom}}, \bibinfo {author} {\bibfnamefont {M.~A.}\ \bibnamefont
  {Scheel}}, \bibinfo {author} {\bibfnamefont {L.~E.}\ \bibnamefont {Kidder}},
  \bibinfo {author} {\bibfnamefont {R.}~\bibnamefont {Owen}},\ and\ \bibinfo
  {author} {\bibfnamefont {O.}~\bibnamefont {Rinne}},\ }\bibfield  {title}
  {\bibinfo {title} {A new generalized harmonic evolution system},\ }\href
  {https://doi.org/10.1088/0264-9381/23/16/S09} {\bibfield  {journal} {\bibinfo
   {journal} {Class. Quantum Grav.}\ }\textbf {\bibinfo {volume} {23}},\
  \bibinfo {pages} {S447} (\bibinfo {year} {2006})},\ \Eprint
  {https://arxiv.org/abs/gr-qc/0512093} {arXiv:gr-qc/0512093} \BibitemShut
  {NoStop}%
\bibitem [{\citenamefont {Szilagyi}\ \emph {et~al.}(2009)\citenamefont
  {Szilagyi}, \citenamefont {Lindblom},\ and\ \citenamefont
  {Scheel}}]{SziLinSch09}%
  \BibitemOpen
  \bibfield  {author} {\bibinfo {author} {\bibfnamefont {B.}~\bibnamefont
  {Szilagyi}}, \bibinfo {author} {\bibfnamefont {L.}~\bibnamefont {Lindblom}},\
  and\ \bibinfo {author} {\bibfnamefont {M.~A.}\ \bibnamefont {Scheel}},\
  }\bibfield  {title} {\bibinfo {title} {Simulations of binary black hole
  mergers using spectral methods},\ }\href
  {https://doi.org/10.1103/PhysRevD.80.124010} {\bibfield  {journal} {\bibinfo
  {journal} {Phys. Rev.}\ }\textbf {\bibinfo {volume} {D80}},\ \bibinfo {pages}
  {124010} (\bibinfo {year} {2009})},\ \Eprint
  {https://arxiv.org/abs/0909.3557} {arXiv:0909.3557 [gr-qc]} \BibitemShut
  {NoStop}%
\bibitem [{\citenamefont {Baumgarte}\ and\ \citenamefont
  {Shapiro}(1998)}]{BauSha98}%
  \BibitemOpen
  \bibfield  {author} {\bibinfo {author} {\bibfnamefont {T.~W.}\ \bibnamefont
  {Baumgarte}}\ and\ \bibinfo {author} {\bibfnamefont {S.~L.}\ \bibnamefont
  {Shapiro}},\ }\bibfield  {title} {\bibinfo {title} {On the {N}umerical
  integration of {E}instein's field equations},\ }\href
  {https://doi.org/10.1103/PhysRevD.59.024007} {\bibfield  {journal} {\bibinfo
  {journal} {Phys. Rev. D}\ }\textbf {\bibinfo {volume} {59}},\ \bibinfo
  {pages} {024007} (\bibinfo {year} {1998})},\ \Eprint
  {https://arxiv.org/abs/gr-qc/9810065} {arXiv:gr-qc/9810065} \BibitemShut
  {NoStop}%
\bibitem [{\citenamefont {Shibata}\ and\ \citenamefont
  {Nakamura}(1995)}]{ShiNak95}%
  \BibitemOpen
  \bibfield  {author} {\bibinfo {author} {\bibfnamefont {M.}~\bibnamefont
  {Shibata}}\ and\ \bibinfo {author} {\bibfnamefont {T.}~\bibnamefont
  {Nakamura}},\ }\bibfield  {title} {\bibinfo {title} {Evolution of
  three-dimensional gravitational waves: {H}armonic slicing case},\ }\href
  {https://doi.org/10.1103/PhysRevD.52.5428} {\bibfield  {journal} {\bibinfo
  {journal} {Phys. Rev. D}\ }\textbf {\bibinfo {volume} {52}},\ \bibinfo
  {pages} {5428} (\bibinfo {year} {1995})}\BibitemShut {NoStop}%
\bibitem [{\citenamefont {Nakamura}\ \emph {et~al.}(1987)\citenamefont
  {Nakamura}, \citenamefont {Oohara},\ and\ \citenamefont
  {Kojima}}]{NakOohKoj87}%
  \BibitemOpen
  \bibfield  {author} {\bibinfo {author} {\bibfnamefont {T.}~\bibnamefont
  {Nakamura}}, \bibinfo {author} {\bibfnamefont {K.}~\bibnamefont {Oohara}},\
  and\ \bibinfo {author} {\bibfnamefont {Y.}~\bibnamefont {Kojima}},\
  }\bibfield  {title} {\bibinfo {title} {General relativistic collapse to black
  holes and gravitational waves from black holes},\ }\href
  {https://doi.org/10.1143/PTPS.90.1} {\bibfield  {journal} {\bibinfo
  {journal} {Prog. Theor. Phys. Suppl.}\ }\textbf {\bibinfo {volume} {90}},\
  \bibinfo {pages} {1} (\bibinfo {year} {1987})}\BibitemShut {NoStop}%
\bibitem [{\citenamefont {Bernuzzi}\ and\ \citenamefont
  {Hilditch}(2010)}]{BerHil09}%
  \BibitemOpen
  \bibfield  {author} {\bibinfo {author} {\bibfnamefont {S.}~\bibnamefont
  {Bernuzzi}}\ and\ \bibinfo {author} {\bibfnamefont {D.}~\bibnamefont
  {Hilditch}},\ }\bibfield  {title} {\bibinfo {title} {Constraint violation in
  free evolution schemes: comparing {BSSNOK} with a conformal decomposition of
  {Z}4},\ }\href {https://doi.org/10.1103/PhysRevD.81.084003} {\bibfield
  {journal} {\bibinfo  {journal} {Phys. Rev. D}\ }\textbf {\bibinfo {volume}
  {81}},\ \bibinfo {pages} {084003} (\bibinfo {year} {2010})},\ \Eprint
  {https://arxiv.org/abs/0912.2920} {arXiv:0912.2920 [gr-qc]} \BibitemShut
  {NoStop}%
\bibitem [{\citenamefont {Alic}\ \emph {et~al.}(2012)\citenamefont {Alic},
  \citenamefont {Bona-Casas}, \citenamefont {Bona}, \citenamefont {Rezzolla},\
  and\ \citenamefont {Palenzuela}}]{AliBonBon11}%
  \BibitemOpen
  \bibfield  {author} {\bibinfo {author} {\bibfnamefont {D.}~\bibnamefont
  {Alic}}, \bibinfo {author} {\bibfnamefont {C.}~\bibnamefont {Bona-Casas}},
  \bibinfo {author} {\bibfnamefont {C.}~\bibnamefont {Bona}}, \bibinfo {author}
  {\bibfnamefont {L.}~\bibnamefont {Rezzolla}},\ and\ \bibinfo {author}
  {\bibfnamefont {C.}~\bibnamefont {Palenzuela}},\ }\bibfield  {title}
  {\bibinfo {title} {{Conformal and covariant formulation of the Z4 system with
  constraint-violation damping}},\ }\href
  {https://doi.org/10.1103/PhysRevD.85.064040} {\bibfield  {journal} {\bibinfo
  {journal} {Phys. Rev. D}\ }\textbf {\bibinfo {volume} {85}},\ \bibinfo
  {pages} {064040} (\bibinfo {year} {2012})},\ \Eprint
  {https://arxiv.org/abs/1106.2254} {arXiv:1106.2254 [gr-qc]} \BibitemShut
  {NoStop}%
\bibitem [{\citenamefont {Hilditch}\ \emph {et~al.}(2013)\citenamefont
  {Hilditch}, \citenamefont {Bernuzzi}, \citenamefont {Thierfelder},
  \citenamefont {Cao}, \citenamefont {Tichy},\ and\ \citenamefont
  {Br{\"u}gmann}}]{HilBerThi12}%
  \BibitemOpen
  \bibfield  {author} {\bibinfo {author} {\bibfnamefont {D.}~\bibnamefont
  {Hilditch}}, \bibinfo {author} {\bibfnamefont {S.}~\bibnamefont {Bernuzzi}},
  \bibinfo {author} {\bibfnamefont {M.}~\bibnamefont {Thierfelder}}, \bibinfo
  {author} {\bibfnamefont {Z.}~\bibnamefont {Cao}}, \bibinfo {author}
  {\bibfnamefont {W.}~\bibnamefont {Tichy}},\ and\ \bibinfo {author}
  {\bibfnamefont {B.}~\bibnamefont {Br{\"u}gmann}},\ }\bibfield  {title}
  {\bibinfo {title} {Compact binary evolutions with the {Z}4c formulation},\
  }\href {https://doi.org/10.1103/PhysRevD.88.084057} {\bibfield  {journal}
  {\bibinfo  {journal} {Phys. Rev. D}\ }\textbf {\bibinfo {volume} {88}},\
  \bibinfo {pages} {084057} (\bibinfo {year} {2013})},\ \Eprint
  {https://arxiv.org/abs/1212.2901} {arXiv:1212.2901 [gr-qc]} \BibitemShut
  {NoStop}%
\bibitem [{\citenamefont {Bona}\ \emph {et~al.}(1995)\citenamefont {Bona},
  \citenamefont {Mass\'o}, \citenamefont {Seidel},\ and\ \citenamefont
  {Stela}}]{BonMasSei94}%
  \BibitemOpen
  \bibfield  {author} {\bibinfo {author} {\bibfnamefont {C.}~\bibnamefont
  {Bona}}, \bibinfo {author} {\bibfnamefont {J.}~\bibnamefont {Mass\'o}},
  \bibinfo {author} {\bibfnamefont {E.}~\bibnamefont {Seidel}},\ and\ \bibinfo
  {author} {\bibfnamefont {J.}~\bibnamefont {Stela}},\ }\bibfield  {title}
  {\bibinfo {title} {New {F}ormalism for {N}umerical {R}elativity},\ }\href
  {https://doi.org/10.1103/PhysRevLett.75.600} {\bibfield  {journal} {\bibinfo
  {journal} {Phys. Rev. Lett.}\ }\textbf {\bibinfo {volume} {75}},\ \bibinfo
  {pages} {600} (\bibinfo {year} {1995})},\ \Eprint
  {https://arxiv.org/abs/gr-qc/9412071} {arXiv:gr-qc/9412071} \BibitemShut
  {NoStop}%
\bibitem [{\citenamefont {Brodbeck}\ \emph {et~al.}(1999)\citenamefont
  {Brodbeck}, \citenamefont {Frittelli}, \citenamefont {H{\"u}bner},\ and\
  \citenamefont {Reula}}]{BroFriHub98}%
  \BibitemOpen
  \bibfield  {author} {\bibinfo {author} {\bibfnamefont {O.}~\bibnamefont
  {Brodbeck}}, \bibinfo {author} {\bibfnamefont {S.}~\bibnamefont {Frittelli}},
  \bibinfo {author} {\bibfnamefont {P.}~\bibnamefont {H{\"u}bner}},\ and\
  \bibinfo {author} {\bibfnamefont {O.~A.}\ \bibnamefont {Reula}},\ }\bibfield
  {title} {\bibinfo {title} {{E}instein's equations with asymptotically stable
  constraint propogation},\ }\href {https://doi.org/10.1063/1.532694}
  {\bibfield  {journal} {\bibinfo  {journal} {J. Math. Phys.}\ }\textbf
  {\bibinfo {volume} {40}},\ \bibinfo {pages} {909} (\bibinfo {year} {1999})},\
  \Eprint {https://arxiv.org/abs/gr-qc/9809023} {arXiv:gr-qc/9809023}
  \BibitemShut {NoStop}%
\bibitem [{\citenamefont {Gundlach}\ \emph {et~al.}(2005)\citenamefont
  {Gundlach}, \citenamefont {Martin-Garcia}, \citenamefont {Calabrese},\ and\
  \citenamefont {Hinder}}]{GunGarCal05}%
  \BibitemOpen
  \bibfield  {author} {\bibinfo {author} {\bibfnamefont {C.}~\bibnamefont
  {Gundlach}}, \bibinfo {author} {\bibfnamefont {J.~M.}\ \bibnamefont
  {Martin-Garcia}}, \bibinfo {author} {\bibfnamefont {G.}~\bibnamefont
  {Calabrese}},\ and\ \bibinfo {author} {\bibfnamefont {I.}~\bibnamefont
  {Hinder}},\ }\bibfield  {title} {\bibinfo {title} {Constraint damping in the
  {Z4} formulation and harmonic gauge},\ }\href
  {https://doi.org/10.1088/0264-9381/22/17/025} {\bibfield  {journal} {\bibinfo
   {journal} {Class. Quantum Grav.}\ }\textbf {\bibinfo {volume} {22}},\
  \bibinfo {pages} {3767} (\bibinfo {year} {2005})},\ \Eprint
  {https://arxiv.org/abs/gr-qc/0504114} {arXiv:gr-qc/0504114} \BibitemShut
  {NoStop}%
\bibitem [{\citenamefont {Bhattacharyya}\ \emph {et~al.}(2021)\citenamefont
  {Bhattacharyya}, \citenamefont {Hilditch}, \citenamefont {Rajesh~Nayak},
  \citenamefont {Renkhoff}, \citenamefont {R\"uter},\ and\ \citenamefont
  {Br\"ugmann}}]{BhaHilRaj21}%
  \BibitemOpen
  \bibfield  {author} {\bibinfo {author} {\bibfnamefont {M.~K.}\ \bibnamefont
  {Bhattacharyya}}, \bibinfo {author} {\bibfnamefont {D.}~\bibnamefont
  {Hilditch}}, \bibinfo {author} {\bibfnamefont {K.}~\bibnamefont
  {Rajesh~Nayak}}, \bibinfo {author} {\bibfnamefont {S.}~\bibnamefont
  {Renkhoff}}, \bibinfo {author} {\bibfnamefont {H.~R.}\ \bibnamefont
  {R\"uter}},\ and\ \bibinfo {author} {\bibfnamefont {B.}~\bibnamefont
  {Br\"ugmann}},\ }\bibfield  {title} {\bibinfo {title} {{Implementation of the
  dual foliation generalized harmonic gauge formulation with application to
  spherical black hole excision}},\ }\href
  {https://doi.org/10.1103/PhysRevD.103.064072} {\bibfield  {journal} {\bibinfo
   {journal} {Phys. Rev. D}\ }\textbf {\bibinfo {volume} {103}},\ \bibinfo
  {pages} {064072} (\bibinfo {year} {2021})},\ \Eprint
  {https://arxiv.org/abs/2101.12094} {arXiv:2101.12094 [gr-qc]} \BibitemShut
  {NoStop}%
\bibitem [{\citenamefont {Alcubierre}(2008)}]{Alc08}%
  \BibitemOpen
  \bibfield  {author} {\bibinfo {author} {\bibfnamefont {M.}~\bibnamefont
  {Alcubierre}},\ }\href@noop {} {\emph {\bibinfo {title} {Introduction to 3+1
  Numerical Relativity}}}\ (\bibinfo  {publisher} {Oxford University Press},\
  \bibinfo {address} {Oxford},\ \bibinfo {year} {2008})\BibitemShut {NoStop}%
\bibitem [{\citenamefont {Cors}(2023)}]{Cor23}%
  \BibitemOpen
  \bibfield  {author} {\bibinfo {author} {\bibfnamefont {D.}~\bibnamefont
  {Cors}},\ }\emph {\bibinfo {title} {Towards Finer Estimations of the
  Threshold of Black Hole Formation}},\ \href@noop {} {Ph.D. thesis},\ \bibinfo
   {school} {Friedrich-Schiller-Universit{\"a}t Jena}, \bibinfo {address}
  {Jena, Germany} (\bibinfo {year} {2023}),\ \bibinfo {note} {in
  preparation.}\BibitemShut {Stop}%
\end{thebibliography}%
\end{document}